\pdfoutput=1
\documentclass[sigconf]{acmart}
\usepackage{amsmath}
\usepackage{graphicx}
\graphicspath{ {./figures/}}
\usepackage{hyperref}
\usepackage{booktabs}
\usepackage{multirow}
\usepackage[utf8]{inputenc}
\usepackage[T2A,T1]{fontenc}
\usepackage{enumitem}
\usepackage{colortbl}
\usepackage{xcolor}
\usepackage{pifont}

\definecolor{mygray}{RGB}{255, 255, 128}
\definecolor{myblue}{RGB}{190,211,222}
\definecolor{mypink}{RGB}{224,192,196}

\definecolor{lightyellow}{RGB}{255, 255, 128} 

\newcommand*{\graysquare}{\textcolor{mygray}{$\blacksquare$}}
\newcommand*{\bluesquare}{\textcolor{myblue}{$\blacksquare$}}
\newcommand*{\pinksquare}{\textcolor{mypink}{$\blacksquare$}}

\newcolumntype{L}[1]{>{\raggedright\let\newline\\\arraybackslash\hspace{0pt}}m{#1}}
\newcolumntype{C}[1]{>{\centering\let\newline\\\arraybackslash\hspace{0pt}}m{#1}}
\newcolumntype{R}[1]{>{\raggedleft\let\newline\\\arraybackslash\hspace{0pt}}m{#1}}

\newcommand{\cmark}{\ding{51}}
\newcommand{\xmark}{\ding{55}}

\def\BibTeX{{\rm B\kern-.05em{\sc i\kern-.025em b}\kern-.08em
    T\kern-.1667em\lower.7ex\hbox{E}\kern-.125emX}}

\copyrightyear{2025}
\acmYear{2025}
\setcopyright{acmlicensed}\acmConference[ASIA CCS '25]{ACM Asia Conference on Computer and Communications Security}{August 25--29, 2025}{Hanoi, Vietnam}
\acmBooktitle{ACM Asia Conference on Computer and Communications Security (ASIA CCS '25), August 25--29, 2025, Hanoi, Vietnam}
\acmDOI{10.1145/3708821.3733861}
\acmISBN{979-8-4007-1410-8/2025/08}
\begin{document}

\title{\textsc{Pitch}: AI-assisted Tagging of Deepfake Audio Calls using Challenge-Response}
\renewcommand{\shorttitle}{PITCH: AI-assisted Tagging of Deepfake Audio Calls}

\author{Govind Mittal}
\email{mittal@nyu.edu}
\affiliation{%
  \institution{New York University}
  \city{Brooklyn}
  \state{NY}
  \country{USA}
}

\author{Arthur Jakobsson}
\email{ajakobss@andrew.cmu.edu}
\affiliation{%
  \institution{Carnegie Mellon University}
  \city{Pittsburgh}
  \state{PA}
  \country{USA}
}

\author{Kelly O. Marshall}
\email{km3888@nyu.edu}
\affiliation{%
  \institution{New York University}
  \city{Brooklyn}
  \state{NY}
  \country{USA}
}

\author{Chinmay Hegde}
\email{chinmay.h@nyu.edu}
\affiliation{%
  \institution{New York University}
  \city{Brooklyn}
  \state{NY}
  \country{USA}
}

\author{Nasir Memon}
\email{memon@nyu.edu}
\affiliation{%
  \institution{New York University}
  \city{Brooklyn}
  \state{NY}
  \country{USA}
}

\renewcommand{\shortauthors}{Mittal et al.}

\begin{abstract}
The rise of AI voice-cloning technology, particularly audio Real-time Deepfakes (RTDFs), has intensified social engineering attacks by enabling real-time voice impersonation that bypasses conventional enrollment-based authentication. This technology represents an existential threat to phone-based authentication systems, while total identity fraud losses reached \$43 billion. Between 2022-23, deepfake attempts occured once every five minutes, with 26\% of Americans reporting they have encountered deepfake scams and 9\% falling victim. Unlike traditional robocalls, these personalized AI-generated voice attacks target high-value accounts and circumvent existing defensive measures, creating an urgent cybersecurity challenge. To address this, we propose \textsc{Pitch}, a robust challenge-response method to detect and tag interactive deepfake audio calls. We developed a comprehensive taxonomy of audio challenges based on the human auditory system, linguistics, and environmental factors, yielding 20 prospective challenges. Testing against leading voice-cloning systems using a novel dataset (18,600 original and 1.6 million deepfake samples from 100 users), \textsc{Pitch}'s challenges enhanced machine detection capabilities to 88.7\% AUROC score, enabling us to identify 10 highly-effective challenges.

For human evaluation, we filtered a challenging, balanced subset on which human evaluators independently achieved 72.6\% accuracy, while machines scored 87.7\%. Recognizing that call environments require human control, we developed a novel human-AI collaborative system that tags suspicious calls as "Deepfake-likely." Contrary to prior findings, we discovered that integrating human intuition with machine precision offers complementary advantages, giving users maximum control while boosting detection accuracy to 84.5\%. This significant improvement situates \textsc{Pitch}'s potential as an AI-assisted pre-screener for verifying calls, offering an adaptable approach to combat real-time voice-cloning attacks while maintaining human decision authority. 
Access code and dataset at \href{https://govindm.me/pitch}{govindm.me/pitch}
\end{abstract}

\maketitle

\section{Introduction}

Recent advancements in AI-generated voice cloning have blurred the line between authentic and fabricated speech \cite{Li2023StyleTTS2T,tortoise_tts,li2023freevc,wu22f_interspeech,valle-microsoft}. This technological leap has particularly significant implications with the emergence of tools capable of generating convincing speech in real-time, fueling AI-generated impersonation scams. Such capabilities expose a critical vulnerability in billions of calls across contact centers, raising an urgent question: \emph{How can one ascertain the authenticity of a caller in an age where synthetic voices are indistinguishable from real ones?}

\begin{figure*}
\centering
\includegraphics[width=2\columnwidth]{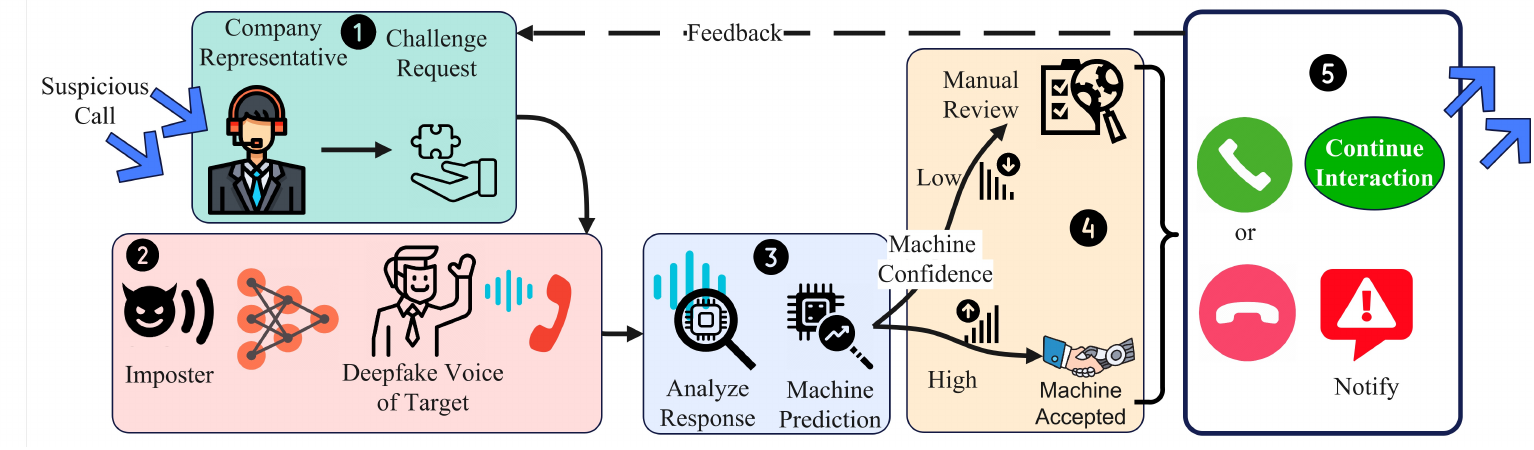}
\caption{Overview of \textsc{Pitch}: (1) A company representative requests an audio challenge from a suspicious caller. (2) The caller attempts the challenge. (3) Machine analysis provides a prediction and confidence level. (4) High-confidence machine predictions are accepted; low-confidence cases are sent for manual review. (5) The final decision is made to either accept the call or reject it and notify the genuine customer. Feedback provides further instructions to the representative.}
\label{fig:overview}
\end{figure*}

\textbf{Motivation.}
The scale of phone scams has reached alarming proportions. In 2022 alone, robocalls in the United States soared to an astonishing 50.5 billion instances, many of them leading to scams. By 2024, scammers had escalated their tactics, using robocalls to impersonate high-profile figures such as U.S. President Biden~\cite{nbcnews_biden_robocall_2024} and the Mayor of New York~\cite{nycmayordeepfakes} for voter suppression and outreach. Truecaller reports that Americans still suffer a \$25B annual loss from phone scams, impacting 21\% of the adult public~\cite{TruecallerSpamScam2024}.

A more sophisticated and targeted form of impersonation fraud leverages \textit{Real-time Deepfakes (RTDFs)}. Scammers utilize speech-to-speech (voice conversion or VC) or text-to-speech (TTS) generation tools to make live deepfake calls that sound convincingly similar to a target's voice. These real-time, interactive calls are more believable and, consequently, more persuasive than traditional scams. Recent data from Entrust shows that deepfake volumes have increased by 3,000\% between 2022-23~\cite{entrust2025}, with attempts now occurring at a rate of one every five minutes.

The human inability to detect such sophisticated voice clones presents a critical vulnerability: only 27\% of people feel confident they could identify if a call from a friend or loved one was real or AI-generated~\cite{mcafee}. These deepfakes represent an unprecedented evolution in social engineering attacks, exploiting deeply-rooted human cognitive biases around voice recognition and trust that traditional security awareness training cannot adequately address. This vulnerability has been exploited in high-profile cases, including a finance worker being convinced to remit \$25 million by an impersonation of their chief financial officer during a group video call~\cite{cnn_deepfake_scam_2024}, a mother being extorted for \$50,000 ransom after hearing what she believed was her daughter's distressed voice~\cite{kidnappergirl}, and an energy company being defrauded of \$243,000 by an impersonation of their boss' voice~\cite{energyscam}. McAfee found that 31\% of respondents had experienced some kind of AI voice scam, believing they were hearing a friend or loved one when they were actually hearing a voice clone~\cite{mcafee}.

While individual scam cases involving the impersonation of loved ones can often be mitigated through precautions such as calling back or asking personal questions, the challenge is far more acute for contact centers of financial institutions, government agencies, and insurance companies. These organizations must continue serving millions of calls daily while protecting their clients' sensitive accounts from increasingly sophisticated impersonators. 

Companies rely on various security protocols to detect impersonators, including speaker verification systems, liveness detection, and knowledge-based authentication. However, a targeted RTDF-based impersonation can potentially bypass all such conventional protections~\cite{kycaml_biometrics_voice_recognition,nytimes_voice_deepfakes,csoonline_generative_ai_kyc}. This represents a fundamentally new attack vector in the cybersecurity landscape, where a compromised voice identity can cascade into broader system infiltration, credential theft, and data breaches that traditional security frameworks are not designed to counteract. The situation is even more dire for organizations that lack the infrastructure to support customers' enrollment in such protective protocols.

While numerous audio deepfake detection methods exist~\cite{yi2023audio}, they are often brittle and primarily designed for static content. These traditional deepfake detectors base their judgment on a limited set of features and assume a non-interactive scenario. As these methods work offline, they are inefficient against deepfake calls where near real-time performance is essential. With deepfakes now accounting for 40\% of all biometric fraud attacks~\cite{entrust2025}, there is an urgent need for robust, real-time approaches to detect and mitigate audio deepfakes in interactive environments.

\textbf{Approach.} In this work, we depart from traditional techniques and leverage challenge-response mechanisms to pursue an alternative approach to detecting deepfake audio calls, which we name \textsc{Pitch}. Challenge-response mechanisms, traditionally pivotal in distinguishing bots from humans via the ubiquitous \textsc{Captcha}, have found new applications with the introduction of \textsc{Gotcha}~\cite{mittal2022gotcha} and \textsc{D-Captcha}~\cite{yasur2023deepfake} for unmasking video and audio deepfakes, respectively. These pioneering studies capitalize on the asymmetric advantage where the burden to maintain high quality in real-time, under challenging situations, is now squarely on the imposter caller. This work aims to harness the \textit{full spectrum} of audio challenges during phone calls and orchestrate a collaborative dynamic between humans and AI, thereby fortifying accuracy and confidence.

We began by curating a detailed taxonomy of audio challenges, categorizing them into eight main types and 22 sub-types, including vocal distortions (e.g., whispering), waveform manipulations (e.g., high-pitch speaking), language-specific articulations (e.g., rolling-R sounds), environmental noise (e.g., clapping while speaking), and background playbacks (e.g., talking over music). We selected and evaluated 20 prospective challenges from the taxonomy.

Our investigation revolves around three research questions: 

\begin{enumerate}[label=\textbf{RQ\arabic*:},nosep]
\item Do challenges enhance machine detection?
\item Can human evaluators harness these challenges to sharpen their discernment?
\item Does supporting humans with automated detectors further improve overall performance?
\end{enumerate}

\textbf{Evaluation.} To address these questions, we collected a novel open-source dataset by engaging 100 participants across mobile and desktop interfaces to produce 18,600 original voice recordings. We subsequently generated 1.6 million samples through state-of-the-art one-shot voice-cloning technologies.

We performed a \textbf{machine-only evaluation} by designing a non-intrusive degradation metric for assessing speech samples on compliance, naturalness, and the preservation of word information. This metric revealed that a select group of 10 challenges dramatically boosted the deepfake detection performance (Area Under ROC or AUROC) to 88.7\%, compared to 56.0\% detecting deepfakes from regular speech, highlighting the benefit of using challenges to aid downstream machine detection. Notably, challenges like whispering, cupping mouth, and playback were among the most effective for machines.

Concurrently, we conducted a \textbf{human-only evaluation} on a \textit{harder balanced} subset of the whole dataset by filtering 6,174 high-quality original and deepfake samples that were also positive matches to their corresponding targets. Humans assessed the compliance and quality of each speech sample and scored an AUROC of 87.7\% and accuracy of 72.6\%, with challenges bringing out discernible degradation in deepfakes.

Informed by these results and the propensity of humans to produce more false positives, we envisioned a framework where \textbf{machines collaborated with humans} in making decisions. This integration makes our approach scalable while keeping the outcome interpretable under human oversight. A subsequent human evaluation with top-performing challenges, using a balanced dataset, revealed that \textit{machine assistance increased human detection accuracy to 79.4\%}. Furthermore, machine assistance significantly boosted human confidence and rectified their errors in 45.4\% of cases (when machines were correct) while causing misjudgments in 29.8\% (when machines were wrong).

Taking this integration a step further, in scenarios \textbf{where machine predictions were highly confident, machines were allowed to take charge and make the final call}. This blend of Human-AI contributions, at a ratio of 56:44, enhanced the overall detection accuracy to 84.5\%, representing a 16.4\% improvement over the human-only baseline (with machine taking full control standing at 87.7\% accuracy). This approach underscores that while humans maintain their primary role in decision-making, strategic machine involvement can dramatically enhance the accuracy of detecting fake calls.

Our findings validated the instrumental role of challenge-response mechanisms in bolstering machine and human capabilities in audio deepfake detection. They showcased a synergy between human insight and AI accuracy that mitigates the threats posed by real-time synthetic speech technologies. As audio is always included during visual interactions, our results can transfer as is to video calls.

We list \textbf{our contributions} as:

\begin{itemize}[nosep,leftmargin=*]
\item A challenge-response system to detect real-time audio deepfakes that significantly advances prior work by providing stronger detection guarantees on a challenging dataset (57\% relative AUC improvement versus 33\%), focusing on making detection interpretable (using human-in-the-loop) \textit{and} scalable (using automated methods), without assuming prior information on the caller.

\item A novel tagging framework, \textsc{Pitch}, that leverages the complementary strengths of human intuition and machine precision through confidence calibration, resulting in a 16.4\% improvement in detection accuracy (84.5\%) over the human-only baseline while maintaining human decision authority.

\item The first systematic and comprehensive taxonomy of speech tasks comprising eight categories and 22 subcategories based on auditory, linguistic, and environmental factors, with extensive evaluation confirming that 10 challenges consistently reveal deepfake vulnerabilities across multiple SoTA voice converters.

\item An open-source novel dataset that dramatically expands the scale of challenge-response evaluation with 18,600 original and 1.6 million deepfake samples (100× larger than prior work), including diverse recording environments, multiple devices, and demographically representative speakers.
\end{itemize}

\vspace{0.35em}
We organized the paper as follows: \S\ref{sec:Background} describes background and the problem, \S\ref{sec:related_work} informs about related work, \S\ref{sec:approach} provides details of the approach, taxonomy, and dataset creation. \S\ref{sec:evaluation} evaluates the RQs, \S\ref{sec:discussion} provides discussion around the solution and \S\ref{sec:conclusion} concludes.

 \section{Background}
\label{sec:Background}

\noindent\textbf{Definition.} An \textit{audio deepfake} refers to a digital impersonation in which an impostor mimics audio characteristics to convincingly match a specific target's likeness. When such impersonations can be executed live with sufficient fidelity, we term them Real-time Deepfakes (RTDFs).

\subsection{Real-time Deepfake Generation}
\noindent\textbf{Text-to-Speech.} Text-to-Speech systems facilitate the conversion of written text into audible speech, enabling automated vocalization of textual content. The process encompasses several stages: textual analysis, phonemic conversion, and speech synthesis. Contemporary TTS methodologies employ various techniques, ranging from concatenative synthesis to advanced neural network-based approaches~\cite{bark_suno_ai, tortoise_tts}. The primary objective of these systems is to generate naturalistic speech output that accurately conveys prosodic and intonational nuances inherent in human speech patterns.
 
\vspace{0.35em}
\noindent\textbf{Voice Conversion.} VC aims to modify the linguistic features of a source speaker's voice to emulate those of a target speaker, while maintaining the original semantic content. This process involves analysing the acoustic properties of \textit{both} source and target vocalizations, followed by the establishment of a feature mapping and subsequent application of this transformation to synthesize modified speech. Recent advancements in the field have focused on real-time processing capabilities~\cite{li2023freevc}, the transfer of emotional vocal characteristics~\cite{zhou2021seen}, and the development of training methodologies utilizing non-parallel datasets~\cite{10243636}. A key challenge in VC research is the simultaneous preservation of speech quality and naturalness while achieving perceptually convincing voice transformations, which we aim to exploit in this work.

\subsection{Problem Description}
\label{sec:Threat Model}

\noindent\textbf{Threat Model.}
Our scenario involves three key parties: an impostor, the intended identity (the target), and a defender. The defender receives a sensitive call and anticipates the target as the caller. However, an impostor could employ an AI-generated speech synthesis tool to communicate with the defender while masquerading as the target. Given the sensitive nature of the information exchange and potential actions, the defender must authenticate the caller's identity before proceeding. Practical examples of this scenario include bank contact centers and job interviews.

\vspace{0.35em}
\noindent\textit{Defenders.} The defender assumes no trust in the potential impostor or their devices. The defender requests that impostors perform specific tasks. While the nature of these requests is public knowledge, the task parameters are randomized at the time of request. Importantly, no identifiable and trusted target information, such as voice biometrics, is collected through an extensive enrollment process. Thus, the defender operates against a stronger threat, independent of helpful identity information.

\vspace{0.35em}
\noindent\textit{Impostors.} An impostor is assumed to have access to sufficient computational resources to run voice-cloning software in real time. They can discern and respond to requests made to them. Additionally, they can obtain a high-quality speech sample of the target, potentially via social media or an unsolicited phone call (e.g., pre-recorded voicemail greetings). This scenario is particularly impactful, as it requires minimal data yet can cause widespread harm.

\vspace{0.35em}
\noindent\textbf{Hypothesis.} Speech communication comprises several vital elements, including phonetics, articulation, pitch, tone, rhythm, stress, voice quality, and fluency, all of which collectively create a distinct auditory experience. We hypothesize that \textit{an audio deepfake system should fail to maintain fidelity in real time while supporting all these speech elements simultaneously}.

\vspace{0.35em}
\noindent\textbf{Distinct Concepts.}
It is important to differentiate our work from these related concepts:

\begin{itemize}[nosep,leftmargin=*]
    \item \textit{Robocalls} are untargeted automated messages distributed en masse to a large population. In contrast, Real-Time Deepfakes (RTDFs), which are the focus of this work, are targeted and possibly curated for the particular receiver.
    
    \item \textit{Liveness detection} mechanisms are, in principle, ineffective against RTDFs, as the impostor generates deepfakes live in real-time. Moreover, we assert that the caller is a person who can converse, making each interaction unfold differently. 
    
    \item Works using \textit{adversarial perturbations} to disrupt the deepfake generation process assume control over the impostor's devices to inject them prior to generation. However, in our threat model, we do not trust the impostor or their devices.
\end{itemize}

\section{Related Work}
\label{sec:related_work}
\noindent\textbf{Deepfake Detection.}
Numerous techniques exist for detecting \textit{static audio deepfakes} based on diverse artifacts, such as vocal tract estimation \cite{blue2022you}, Mel-frequency cepstral coefficients (MFCC) \cite{altalahin2023unmasking}, non-semantic features~\cite{safeear} and vocoder fingerprints \cite{10446798}. 

However, we assert that \textsc{\textsc{Pitch}} is \textit{not a traditional ML-based deepfake detector}; rather, it is a framework that can enhance the performance of current deepfake detectors by amplifying unnatural artifacts via carefully designed challenges.

Through challenge-response mechanisms, \textsc{\textsc{Pitch}} places the burden of proof squarely on the attacker. For instance, if a particular deepfake detector responds well to deepfake artifacts that sound atonal or robotic, \textsc{\textsc{Pitch}} can deploy a tonal challenge (such as ``sounding happy'') against a suspected user to amplify this artifact.

\vspace{0.35em}
\noindent\textbf{Challenge-Response Systems.}
Authentication schemes utilizing challenge-response mechanisms are exemplified by the ubiquitous \textsc{Captcha}, which differentiates bots masquerading as humans on the internet. Audio-\textsc{Captcha} \cite{fanelle2020blind} extends this approach for people with visual impairments. The \textsc{Gotcha} system \cite{mittal2022gotcha} introduced a challenge-response approach for real-time video-only deepfake detection, analyzing eight video challenges. This study focused on interpretability and concluded that challenges lead to consistent and measurable degradation in detecting real-time deepfakes in muted videos.
\begin{table}[t!]
\caption{Comparison of \textsc{Pitch} with Related Challenge-Response Frameworks.}

{\footnotesize
\begin{tabular}{L{1.15cm}C{1.45cm}C{1.45cm}C{1.4cm}C{1.3cm}}
\toprule
\textbf{Feature} & \textbf{\textsc{Pitch} (Ours)} & \textbf{D-CAPTCHA \cite{yasur2023deepfake}} & \textbf{D-CAPTCHA++ \cite{dcaptchaplusplus}} & \textbf{GOTCHA \cite{mittal2022gotcha}} \\
\midrule
Modality & Audio & Audio & Audio & Video \\
\midrule
Key Innovation & Human-AI collaboration + confidence calibration & Challenge-based audio deepfake detection & Adversarial defense for D-CAPTCHA & Challenge-based video deepfake detection \\
\midrule
Challenge Taxonomy & Extensive (8 categories, \newline 22 sub-categories) & Limited \hspace{2em} (no formal taxonomy) & Extends D-CAPTCHA & Facial Video-focused \\
\midrule
Number of Challenges & 20 evaluated, 10 effective & 9 evaluated, 3 effective & Built on D-CAPTCHA & 8 video challenges \\
\midrule
Dataset Size & 18,600 original 1.6M deepfake 100 speakers & 2,500 original 16K deepfake 20 speakers & Based on D-CAPTCHA & Video only \\
\midrule
Environment & Desktop \& mobile, In-the-wild & Controlled laboratory & Controlled laboratory & Controlled laboratory \\
\midrule
Deepfake Models Evaluated & FREE-VC StarGANv2-VC PPG-VC & StarGANv2-VC & StarGANv2-VC & N/A \\
\midrule
Human-AI Integration & Collaborative \& Calibration & None & None & Human-focused \\
\midrule
Detection Performance & 88.7\% AUC up from 56.0\% (harder)
& 95\% Acc. up from 71\% (easier) & Enhanced adversarial robustness & 88\% Human, 80\% Machine AUC \\

\bottomrule
\end{tabular}}
\label{tab:related_comparison}
\end{table}

D-\textsc{Captcha} \cite{yasur2023deepfake} developed a challenge-response system for audio calls, evaluating nine challenges and measuring detection performance based on realism, compliance, and identity against deepfake generators, with StarGANv2-VC \cite{starganv2vc} being the most potent deepfake generator tested. Their research provides a vital proof-of-concept for a more systematic and robust approach, indicating the promising potential of challenge-response systems in detecting deepfake audio calls. D-\textsc{Captcha}++~\cite{dcaptchaplusplus} is a follow-up that proposed an imperceptible adversarial attack on D-\textsc{Captcha} \cite{yasur2023deepfake} and employs projected gradient descent-based adversarial training to defend against it. Our work expands upon this foundation in several key ways as listed in Table~\ref{tab:related_comparison}, specifically by using our comprehensive taxonomy of audio challenges supporting human-AI collaborative integration on in-the-wild evaluations.

\vspace{0.35em}
\noindent\textbf{Comparison with Existing Authentication Protocols.}
Unlike traditional authentication methods (speaker verification, liveness detection, MFA, OTP), \textsc{Pitch} takes an active challenge-response approach to voice verification that addresses fundamental limitations in existing protocols. 

Current speaker verification systems~\cite{liu2024generalizing} require prior enrollment and struggle with natural voice changes over time, requiring regular re-enrollment and maintaining a secure infrastructure to maintain biometric information. Liveness detection mechanisms~\cite{liu2018learning} are, in principle, ineffective against RTDFs, as the deepfake is already live and would be bypass liveness checks. Multi-factor authentication (MFA) introduces significant user friction and deployment complexities across communication channels, while one-time password (OTP) systems depend on secondary channels that are vulnerable to delays and real-time phishing during calls.

In contrast, \textsc{Pitch} requires no prior enrollment, trivially checks for liveness, operates entirely in-band during the call, while actively challenges potential impostors to perform tasks that exploit the fundamental limitations of speech synthesis technology. With 88.7\% detection accuracy against SoTA deepfakes and a human-machine collaborative framework, \textsc{Pitch} provides a complementary security layer that offers superior interpretability through its challenge-response format. Rather than replacing existing authentication infrastructure, \textsc{Pitch} specifically addresses the emerging threat of real-time deepfakes while maintaining compatibility with other verification methods.

\vspace{0.35em}
\noindent\textbf{Human Evaluation of Deepfakes.}
Muller et al. \cite{muller2022human} conducted a study with participants competing against an AI model (RawNet2 \cite{rawnet2}) to identify audio deepfakes. Participants were shown the AI's classification after listening to an audio clip and deciding its authenticity. The study compared human performance against AI in detecting deepfakes without allowing participants to revise their choices. The conclusion was that humans and AI exhibit similar strengths and weaknesses across various spoofing tasks from ASVSpoof2019 \cite{nautsch2021asvspoof}. Contrary to their findings, our work provides empirical evidence that human and AI performances are not strongly correlated, often complementing each other's decision-making.

 \section{Methodology}
\label{sec:approach}

\subsection{Audio Challenges and their Taxonomy}
\label{sec:ChallengeClassification}
We define an \textit{audio challenge} as a task that is:
\begin{itemize}[nosep,leftmargin=*]
    \item plausible to perform during a phone call,
    \item capable of inducing degradation in real-time deepfakes,
    \item supports randomization, and
    \item optionally verifiable by humans.
\end{itemize}

\vspace{0.35em}
Table~\ref{sec:tab:taxonomy} delineates our taxonomy of audio challenges, designed to target fundamental limitations in current deepfake technology: (1) physiological modeling gaps in speech production, (2) limited training diversity in acoustic conditions, (3) inadequate handling of multi-source audio, (4) poor temporal adaptability in real-time processing, and (5) limited capability to maintain voice characteristics across diverse speech modes. The following describes each category and sub-category, with emphasis on how these challenges exploit specific weaknesses in deepfakes.
\begin{table}[t!]
  \caption{Taxonomy of Audio Challenges encompassing eight categories and 22 subcategories with examples. We evaluated the 20 numbered examples as prospective challenges.}
  \centering
  {\footnotesize
  \begin{tabular}{L{1.8cm}L{1.8cm}L{3.6cm}}
\toprule
\textbf{Category} & \textbf{Sub-category} & \textbf{Examples (Selection Index)} \\
\midrule
No Challenge & Read Normally & Regular Speech \textbf{(\#0)} \\
\midrule
\rowcolor{mygray}
 & Vocal Peripherals & Static Mouth \textbf{(\#1)}, Cup Mouth \textbf{(\#2)} \\ 
\rowcolor{mygray}  \multirow{-2}{1.8cm}{Vocal Distortions}
 & Whisper & Whispering \textbf{(\#3)} \\ 
\rowcolor{mygray}  
 & Vocal Cavity & Hold Nose \textbf{(\#4)}, humming \\
\midrule
\rowcolor{mygray}   
 & Frequency & High \textbf{(\#5)}, Low pitch \textbf{(\#6)}, Sing \textbf{(\#7)}\\
\rowcolor{mygray}\multirow{1}{1.8cm}{Waveform}
 & Amplitude & Speak Loudly \textbf{(\#8)} \& Softly \textbf{(\#9)}\\
\rowcolor{mygray}
 & Temporal & Speak Quickly \textbf{(\#10)} \& Slowly \textbf{(\#11)}\\
\midrule
\rowcolor{mygray}   
 & Hesitation & Foreign Words \textbf{(\#12)}, Reverse Count\\
\rowcolor{mygray}   
 & Mimicry & Mimic another Accent \textbf{(\#13)}\\
\rowcolor{mygray}\multirow{-2}{1.8cm}{Language / Articulation}
 & Phonetics & Rolled R's \& Tongue clicks\\
\rowcolor{mygray}   
 & Deception & Sudden Interruption while speaking \\
\midrule
\rowcolor{mygray}   
 & Emotion & Sound happy / sad \textbf{(\#14)} \\
\rowcolor{mygray} \multirow{-2}{1.8cm}{Tone of Voice}
 & Phonology & Questions (inflection) \textbf{(\#15)} \\
\midrule
\rowcolor{myblue}   
 & Vocal & Cough/whistle \textbf{(\#16)} \\
\rowcolor{myblue}   
 & Non-vocal & Clap \textbf{(\#17)}, flick microphone \\
\rowcolor{myblue}\multirow{-2.6}{1.8cm}{Noise}
 & Environmental & Birds, Cars \\
\rowcolor{myblue} 
 & Reverb & Two Mics on same call \\
\midrule
\rowcolor{mypink}   
& Speech & Cross-talk \textbf{(\#18)}\\
\rowcolor{mypink}  \multirow{-2}{1.8cm}{Playback (desktop only)} 
 & Music & Instrumental \textbf{(\#19)}, Lyrical \textbf{(\#20)} \\
\midrule
\multicolumn{3}{c}{\textbf{Out-of-Scope}} \\
\midrule
\rowcolor{myblue}   
 & Unique Habits & Mannerisms, Person-of-Interest \\
\rowcolor{myblue}\multirow{-2}{1.8cm}{Behavioral}
 & Biometric & Voice Print Detection \\
\midrule
\rowcolor{mypink}   
 & Perturbations & Adversarial Perturbations \\
\rowcolor{mypink}\multirow{-2}{1.8cm}{Passive Distortions}
 & Software Editing & Modulating Pitch, Noise, or Bass \\
\bottomrule
\multicolumn{3}{l}{\graysquare Lingual challenge, \bluesquare Non-lingual challenge, \pinksquare Replay challenge} \\
\end{tabular}
}
  \label{sec:tab:taxonomy}
\end{table}

\subsection*{Lingual Challenges}
\label{sec:LingualChallenges}
Lingual challenges prompt individuals to alter their vocal characteristics in distinctive ways, exploiting the degrees of freedom in the human vocal system that are inadequately modeled by deepfakes.

\subsubsection{Vocal Distortions}
These challenges manipulate mouth peripherals and vocal chord resonance, targeting the deepfake weakness in physiological speech modeling:

\begin{itemize}[nosep,leftmargin=*]
\item \textit{Vocal Peripherals:} Impeding the movement of teeth, lips, or tongue (e.g., speaking with a pen between teeth) restricts articulation. This exploits deepfakes' inability to model the physical constraints of the human vocal tract, as current models lack explicit representation of articulatory dynamics \cite{Zhang2022}.

\item \textit{Whisper:} Whispering, characterized by minimal vocal cord vibrations, creates a toneless output with significant spectral differences from normal speech. Deepfake models typically trained on normal speech samples lack sufficient exposure to this mode of vocalization, resulting in poor reconstruction quality.

\item \textit{Vocal Cavity:} Modifying the vocal tract cavity (e.g., speaking while holding one's nose) impacts sound resonance in ways that deepfake generators cannot accurately reproduce, as they do not incorporate acoustic principles of nasal and oral interaction.
\end{itemize}

\vspace{0.35em}
\noindent\textbf{Waveform Challenges.}
These involve modifying key speech characteristics in the sound waveform, exploiting the limited spectral and temporal diversity in deepfake training data:

\begin{itemize}[nosep,leftmargin=*]
\item \textit{Frequency:} Varying voice pitch, from unusually low or high pitches to singing melodies, tests the waveform diversity an RTDF can handle. Most deepfake systems are trained on datasets with limited frequency range (e.g., VCTK corpus), resulting in degraded performance for outlier frequencies \cite{singing2022icassp}.

\item \textit{Amplitude:} Modulating speaking volume introduces variations in signal-to-noise ratio and harmonic structure that challenge the dynamic range capabilities of voice conversion systems, which are typically optimized for conversational amplitude levels.

\item \textit{Temporal:} Abnormal speech rates stress deepfakes' temporal modeling capabilities. Current systems struggle with maintaining linguistic coherence under extreme temporal constraints due to limitations in their sequence modeling architecture.
\end{itemize}

\vspace{0.35em}
\noindent\textbf{Language and Articulation Challenges.}
These involve articulating uncommon language patterns that exploit the limited linguistic diversity in training corpora:

\begin{itemize}[nosep,leftmargin=*]
\item \textit{Hesitation:} Pronouncing difficult sequences or foreign words tests RTDFs' ability to handle `guessing' phonetic patterns, exposing their inability to model uncertainty in human conversations.

\item \textit{Mimicry:} Users mimicking known accents may confound RTDFs as they typically assume consistent source speaker characteristics rather than intentional style manipulations.

\item \textit{Phonetics:} Reproducing sounds not native to the speaker's language (e.g., rolling Rs) challenges RTDFs' adaptability to phonological structures absent in their training data.

\item \textit{Deception:} Unexpected commands to cease speaking test reaction times, exposing RTDFs' lack of human-like responsiveness and inability to model pragmatic aspects of communication.
\end{itemize}

\vspace{0.35em}
\noindent\textbf{Tone of Voice.}
\begin{itemize}[nosep,leftmargin=*]
\item \textit{Emotion:} Speaking with discernible emotions tests RTDFs' ability to replicate nuanced expressions. Most voice converters focus primarily on content and speaker identity preservation, with emotional prosody often lost or flattened during conversion.

\item \textit{Phonology:} Altering inflection, particularly at sentence endings, challenges RTDFs' prosodic capabilities, as current models inadequately capture the full complexity of human intonation patterns, especially those that convey semantic nuance.
\end{itemize}

\subsection*{Non-lingual Challenges}
These challenges rely on human-generated noises, environmental sounds, or unique speech patterns that exploit deepfakes' limited ability to process complex acoustic environments.

\vspace{0.35em}
\noindent\textbf{Noise Challenges}
Inducing sounds from the speaker or environment to create artifacts in deepfakes:

\begin{itemize}[nosep,leftmargin=*]
\item \textit{Vocal Noises:} Includes coughs, whistles, and loud breathing. These non-speech vocalizations occupy spectral regions poorly modeled by deepfake systems, which are primarily optimized for linguistic content.

\item \textit{Non-vocal Noises:} Involves producing intentional sounds like clapping, snapping fingers, or microphone manipulation. These create transient acoustic events that challenge voice conversion systems' ability to maintain coherent voice characteristics during hybrid audio scenarios.

\item \textit{Environmental Noises:} Incorporates background sounds like bird songs or traffic noise. Voice converters typically struggle with source separation when faced with complex acoustic environments, often degrading both the background and speech.

\item \textit{Reverb:} Creates feedback loops or reverberations during the call, exploiting the limited ability of deepfakes to model acoustic environment properties and their effect on speech propagation.
\end{itemize}

\vspace{0.35em}
\noindent\textbf{Behavioral Challenges. (out-of-scope)}
These utilize known attributes of an individual's speaking style or biometrics but are not explored in this work due to our assumption of having no prior identity information.

\subsection*{Replay Challenges}

\vspace{0.35em}
\noindent\textbf{Playback.}
This category involves playing recordings that expose deepfakes' weakness in handling multi-source audio:
\begin{itemize}[nosep,leftmargin=*]
\item \textit{Speech:} Subjects speak while a recorded voice is playing, creating a multi-speaker scenario that forces voice conversion models to simultaneously process overlapping linguistic content, a capability not well-developed in current systems.

\item \textit{Music:} Speaking over music introduces non-speech audio patterns with harmonic structures that interfere with the voice conversion process. Current deepfake systems lack robust source separation capabilities needed to cleanly process speech in such complex acoustic environments.
\end{itemize}

\vspace{0.35em}
\noindent\textbf{Passive Distortions (out-of-scope):} These involve manipulating the caller's voice or playing adversarial audio and are considered out-of-scope as they assume trust in the caller's equipment.

\vspace{0.35em}
By systematically exploiting these weaknesses, our challenges provide an effective framework for detecting audio deepfakes in real-world scenarios. To corroborate this, we begin by curating a data of audio challenges.

\subsection{Challenges Dataset Curation}
\label{sec:data_collection}
\noindent\textbf{Ethics and Recruitment.} Following our Institutional Review Board (IRB) approval, we recruited participants using Prolific. We informed all subjects about the intended use of their data for creating and evaluating deepfakes. Participants provided consent for video and audio recording during task performance and agreed to share their data for future secondary research. 
 
We carefully considered the ethical implications of creating and using deepfakes, including potential misuse and privacy concerns. To mitigate these risks, we implemented strict data handling protocols and enforcement to access the generated deepfakes. Our protocols included an Institutional Data Use Agreement restricting usage to approved research purposes, using online storage designed specifically for high-risk data and a participant withdrawal mechanism enabling automated data purging upon request. 

\vspace{0.35em}
\noindent\textbf{Study Design.} Our study encompassed recordings in both desktop and smartphone environments. In the \textit{desktop} setting, we explored 21 tasks, including standard speech in challenge \#0, i.e. the "no-challenge" condition. For 20 of the 21 challenges, each participant read ten phonetically diverse sentences per challenge (see Table~\ref{tab:sentences}). We chose ten sentences per challenge to balance comprehensive data collection with participant fatigue. We limited challenges involving coughing or whistling to one instance. Consequently, each successful desktop participant recorded 201 samples.

We designed the \textit{smartphone} study to simulate a mobile user scenario with potentially limited task-performance freedom. As such, the smartphone dataset constituted a strict subset of the desktop dataset. We excluded playback challenges (\#18, \#19, \#20), which necessitated a secondary device for background sound playback—an unrealistic assumption in a mobile context. Thus, each successful smartphone participant recorded 171 samples.

We developed the experiments using the Gorilla Experiment Builder (see Fig~\ref{fig:data_collection}). The initial data collection involved 100 subjects, equally distributed between smartphone and desktop environments, yielding \textit{18,600 original recordings}. This dataset is comparable in scale to the well-known VCTK speech dataset \cite{vctk}, which contains 110 speakers with 400 samples each (totaling 44,000 recordings), while offering \textit{significantly greater diversity}. Our participant pool consisted of individuals aged 18-65, near equal gender distribution, and is multi-racial. This diverse demographic helped ensure a representative sample of speech patterns and characteristics. \footnote{We open source our instrument at \url{https://app.gorilla.sc/openmaterials/722500}}

\vspace{0.35em}
\noindent\textbf{Data Pre-processing.} We extracted the 48kHz audio stream from the recorded samples (.webm) and down-sampled them to 16kHz lossless audio (.wav) files for downstream tasks. We employed WhisperX \cite{bain2022whisperx} to transcribe the down-sampled audio samples and obtain timestamps for speech segments. We then concatenated the segmented parts to clip the speech's start and end, effectively removing silent intervals.

\vspace{0.35em}
\noindent\textbf{Manual Quality Check of Original Recordings:} To ensure dataset quality, we engaged 45 human evaluators to verify the presence of required challenges in the pre-processed audio recordings. To optimize the evaluation process, we randomly sampled one audio recording per subject per challenge for compliance checking, assuming consistent performance across multiple attempts.

Three English-speaking evaluators—a native monolingual, a non-native multilingual, and a tie-breaker—verified each sample. \textit{This approach aimed to enhance inclusivity and mitigate demographic bias}. We made compliance decisions using majority voting, resulting in an 84\% mean compliance across 100 subjects and 89\% across challenges. 

\vspace{0.35em}
\noindent\textbf{Choosing RTDF generators.} Our analysis revealed that SoTA Text-to-Speech (TTS) models such as TorToise \cite{tortoise_tts} and Bark \cite{bark_suno_ai} excelled in content modulation but fell short in capturing speech nuances like emotions or prosody, significantly limiting their effectiveness against our challenges. \textit{Consequently, we shifted our focus to speech-to-speech systems}, which convert source speech into the targeted speaker's voice while preserving subtle speech elements.

For our evaluation, we selected FREE-VC~\cite{li2023freevc} as the primary deepfake generator after assessment of available voice conversion technologies. FREE-VC represents current state-of-the-art in one-shot voice conversion, achieving \textbf{superior quality} (MOS 3.92±0.14 versus StarGANv2-VC's 3.49±0.11) while requiring only 3-5 seconds of target audio. This \textbf{aligns with realistic threat scenarios} where attackers have limited access to victim voice samples. Its \textbf{computational efficiency} (98ms processing per second of audio) enables real-time conversion necessary for interactive call scenarios, and independent security evaluations ~\cite{korshunov2023vulnerability} confirm it presents a "high threat" to speaker verification systems with a 78.3\% success rate in bypassing authentication (compared to 64.7\% for alternatives). As an open-source implementation, FREE-VC represents an \textbf{accessible tool for potential attackers}.

By evaluating our system against this sophisticated model, we ensure challenges effective against FREE-VC will likely work against less advanced voice conversion technologies.  However, to prevent relying only on one VC, we had also evaluated StarGANv2-VC~\cite{starganv2vc} and PPG-VC~\cite{ppgvc} (see Table~\ref{tab:rebuttal_table}).

\vspace{0.35em}
\noindent\textbf{Generating Deepfakes:} For deepfake generation, we considered each of our 100 participants as a target, with all others serving as imposters, resulting in $100\times99=9,900$ unique target-imposter pairs. Our challenges utilized one-sentence ($4.5\pm1.1$s) samples of both target and imposter speech. This setup intentionally varied speech content between target and imposter to reflect realistic constraints. We converted all collected samples into deepfakes using FreeVC \cite{li2023freevc}, generating a total of 1.6 million samples (99 imposters $\times$ 18,600 samples, minus 241,000 failure cases).

With the challenges selected and speech dataset of original and deepfakes in place, we evaluate them next with machines, humans and subsequently their collaboration.

\section{Evaluation}
\label{sec:evaluation}
An impersonator's goal is to mimic a target individual during a call:
\begin{enumerate}[nosep,leftmargin=*]
\item By adhering to the challenge in terms of task and content,
\item Maintaining the target speaker's identity, and
\item Achieving a high level of realism.
\end{enumerate}

Conversely, our challenges aim to disrupt them.

\subsection{RQ1. Machine Evaluation}
\label{sec:machine_evaluation}

In this subsection, we evaluate the effectiveness of challenges against voice-cloning attacks using an automated authentication system. Our machine evaluation approach is designed to comprehensively assess the effectiveness of challenges against deepfake attempts without requiring reference samples or human interventions. We selected three non-intrusive metrics based on key aspects that determine successful voice impersonation: (1) Challenge Compliance measures the ability to perform the specific requested task, (2) Content Compliance quantifies information preservation in the delivered message, and (3) Realism assesses the naturalness of speech production. These metrics collectively provide a multi-dimensional assessment that captures both technical and perceptual aspects of deepfake quality.

\vspace{0.35em}\noindent\textbf{Challenge Compliance $\mathcal{C}$.} We assess this criterion using a set of 18 binary classifiers, trained on the original speech data described in \S\ref{sec:data_collection}. We designed each classifier to distinguish between a specific challenge and regular speech (challenge \#0). We partitioned the dataset by subject identity for 5-fold cross-validation, using 90 identities and reserving 10 for a final test set.

We based each classifier on a fine-tuned Wav2Vec2 model \cite{wav2vec2020}. We chose Wav2Vec2 as the feature backbone, as it was pretrained on 50,000 hours of unlabeled speech and provided superior performance against MFCC, LFCC or Mel Spectrogram-based classifiers. 

We fine-tuned each Wav2Vec2 model for 30 epochs with a batch size of 64 and a learning rate of 5e-5. During training, we monitored validation accuracy after each epoch, retaining the best-performing model. This process yielded mean validation and test accuracy of 87.5\% and 83.6\% $\pm$ 8.0\%, respectively, and achieved a mean AUROC of 93.1\%. For context, human evaluators rated compliance at 86.0\% (provided by human evaluators to the original challenges). Thus, our models are close to ground truth accuracy. For details on Wav2Vec2 and further challenge compliance detectors, refer~\S\ref{sec:wav2vec2_architecture}.

For the remaining 2 out of 20 challenges—speaking foreign words (challenge \#12) and asking questions (challenge \#15)—we assessed compliance through speech transcription (see Table~\ref{tab:detector_scores}).

While we follow machine learning protocols for dataset splitting, we apply the compliance detectors to all identities. Importantly, we \textit{do not use any deepfakes} for training or validation, as we are only testing compliance to the challenge.

\vspace{0.35em}\noindent\textbf{Content Compliance.} We employ Word Information Lost (WIL) \cite{word-info-lost} to quantify the discrepancy between (a) transcribing the audio sample using WhisperX and (b) the original script. WIL was chosen over standard Word Error Rate (WER) for its emphasis on keywords and de-emphasis of fillers, better reflecting natural conversation dynamics. WIL values range from 0 (perfect preservation) to 1 (complete information loss). This metric is applied to all challenges except those involving non-verbal sounds (e.g., cough/whistle). 

\vspace{0.35em}\noindent\textbf{Realism $\mathcal{R}$.} We utilize the Non-Intrusive Speech Quality Assessment (NISQA) \cite{NISQA} to evaluate realism. \textit{NISQA} is an ML framework for speech quality prediction, which was trained and evaluated on 81 diverse datasets. It has been extensively adopted and recognized by the Speech community. NISQA predicts a 5-point mean opinion score (pMOS) for a given audio sample without requiring a reference signal, aligning with general speech research applications. 
\begin{figure*}[h!]
\centering
\includegraphics[width=2\columnwidth]{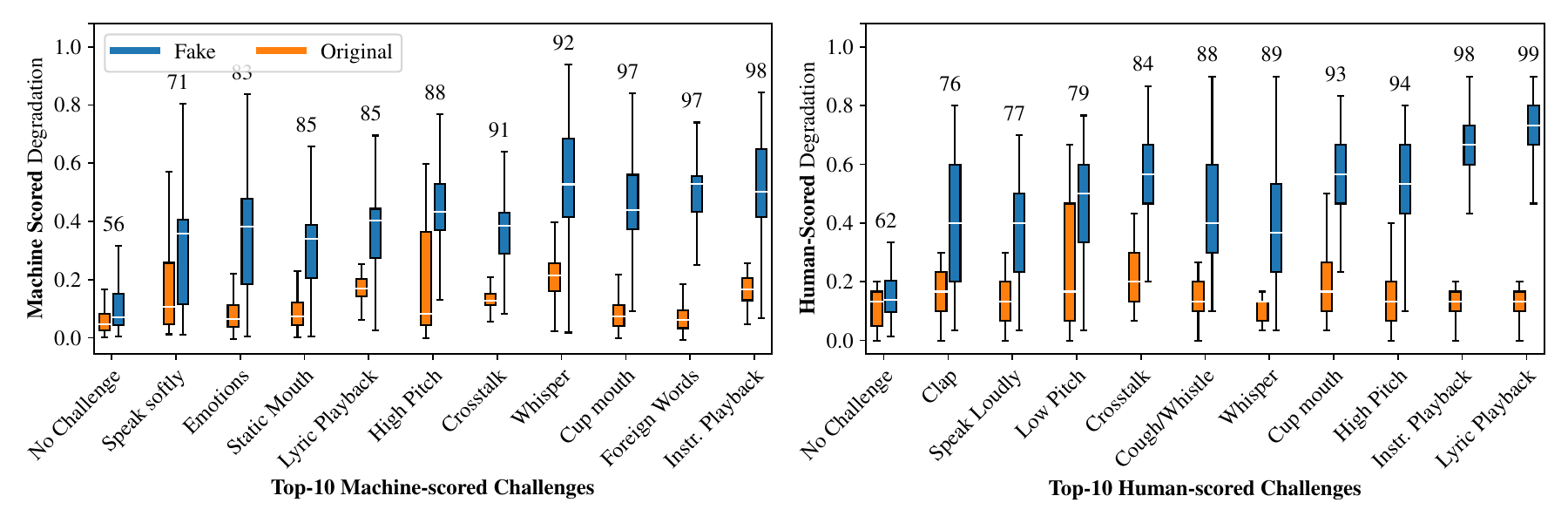}
\caption{\small Comparison of degradation scores between machine (left) and human (right) evaluations across top-10 challenges. Both panels display boxplots for fake (blue) and original (orange) audio samples with AUROC percentages shown above. Left: Machine-scored degradation arranged by increasing median values of fake samples. Right: Human-scored degradation arranged by increasing AUROC. Higher scores ($\uparrow$) indicate greater degradation and better challenge performance. The different ordering between panels reveals complementary strengths between machine and human detection capabilities. Full version: Fig. \ref{fig:human_boxplot}.}
\label{fig:automated_boxplot}
\end{figure*}

Based on these three measures, we define a machine-scored degradation $M \in [0, 1]$ for an audio sample $(x)$ and challenge $(chal)$ as follows:
\begin{equation*}
M_{chal}(x) = \frac{1}{3} \left[ (1 - C_{chal}(x)) + WIL(x) + \left(1 - \frac{R(x)}{5}\right) \right]
\end{equation*}
\begin{itemize}[nosep,leftmargin=*]
    \item $C_{chal}(x)$ is the binary challenge compliance score (1 if compliant, 0 if non-compliant).
    \item $WIL(x)$ is the Word Information Lost score (higher values indicate more information loss).
    \item $R(x)$ is the NISQA realism score on a scale of 1-5. The division by 5 normalizes the realism score to a 0-1 scale.
\end{itemize}
This balanced formula gives equal weight to all three aspects of deepfake quality assessment. A high degradation score $M$ indicates higher degradation (poor performance of deepfake and good performance of challenge) in at least one dimension, which aligns with our objective of identifying challenges that disrupt deepfakes.

These three components represent complementary but distinct dimensions of deepfake performance. Challenge compliance captures whether the deepfake can adapt to specific vocal tasks; content compliance measures whether semantic meaning is preserved; and realism assesses the natural quality of the speech production. While a high-quality deepfake would need to score well in all three dimensions, our findings reveal that different challenges target different vulnerabilities: some primarily affect compliance (e.g., cup mouth), others mainly impact content preservation (e.g., foreign words), and some primarily degrade realism (e.g., whisper). 

\begin{table}[t!]
    \caption{\small Challenge effectiveness measured by degradation scores. For each challenge, we report: mean degradation scores $\mathcal{M}$ for fake (F) and original (O) samples; corresponding AUC values; and $\Delta$ metrics showing differences from baseline in challenge compliance ($\Delta$C), realism ($\Delta$R), and word information loss ($\Delta$WIL). All values are in percentages, with higher being better for detection. Gray-shaded rows indicate underperforming challenges (AUROC $\leq 70\%$).}
    \centering
    {\footnotesize
    \begin{tabular}{cL{1.58cm}C{0.5cm}C{0.5cm}C{0.68cm}|C{0.72cm}C{0.75cm}C{0.55cm}}
\toprule
\textbf{No.} & \textbf{Challenge} & $\Delta\mathcal{C}$ & $\Delta\mathcal{R}$ & $\Delta$WIL & $\mathcal{M}(F)$  & $\mathcal{M}(O)$ & \textbf{AUC} \\
\midrule
\#0 & No Challenge & - & 8 & 11 & 11 & 5 & 56 \\
\midrule
\#1 &Static Mouth & 22 & 10 & 57 & 31 & 11 &  85 \\
\#2 &Cup mouth & 87 & 11 & 38 & 46 & 9 &  97 \\
\#3 &Whisper & 61 & 27 & 63 & 53 & 22 &  92 \\
\#4 & \cellcolor[gray]{0.6} Hold nose & 55 & 11 & 32 & 29 & 38 & \cellcolor[gray]{0.6} 63 \\
\#5 &High Pitch & 63 & 9 & 28 & 44 & 17 &  86 \\
\#6 & \cellcolor[gray]{0.6} Low Pitch & 35 & 10 & 20 & 25 & 15 & \cellcolor[gray]{0.6} 65 \\
\#7 & \cellcolor[gray]{0.6} Sing & 18 & 10 & 32 & 21 & 9 & \cellcolor[gray]{0.6} 70 \\
\#8 & \cellcolor[gray]{0.6} Speak Loudly & 34 & 7 & 13 & 26 & 16 & \cellcolor[gray]{0.6} 62 \\
\#9 &Speak softly & 36 & 12 & 22 & 29 & 16 &  71 \\
\#10 &\cellcolor[gray]{0.6}  Read quickly & 12 & 8 & 35 & 24 & 14 & \cellcolor[gray]{0.6} 67 \\
\#11 &\cellcolor[gray]{0.6}  Read Slowly & 19 & 11 & 13 & 18 & 20 & \cellcolor[gray]{0.6} 55 \\
\#12 & Foreign Words & - & 10 & 82 & 47 & 6 &  97 \\
\#13 &\cellcolor[gray]{0.6}  Accent & 25 & 9 & 26 & 21 & 18 & \cellcolor[gray]{0.6} 57 \\
\#14 & Emotions & 30 & 9 & 25 & 35 & 10 &  82 \\
\#16 &\cellcolor[gray]{0.6}  Cough/Whistle & 36 & 26 & - & 38 & 34 & \cellcolor[gray]{0.6} 54 \\
\#17 &\cellcolor[gray]{0.6}  Clap & 12 & 24 & 33 & 22 & 19 & \cellcolor[gray]{0.6} 51 \\
\#18 & Crosstalk & 27 & 17 & 65 & 38 & 13 &  90 \\
\#19 & Instr. Playback & 96 & 24 & 40 & 53 & 17 &  98 \\
\#20 & Lyric Playback & 26 & 25 & 61 & 38 & 17 &  85 \\
\midrule
\multicolumn{2}{l}{Average} & 37.3 & 13.6 & 35.8 & 31.7 & \textbf{15.9} & 74.4 \\
\multicolumn{2}{l}{Average (top-10)} & \textbf{49.5} & \textbf{15.4}& \textbf{48.0} & \textbf{39.5}& 13.8 & \textbf{88.7}\\
\bottomrule
\end{tabular} }
    \label{tab:automated_evaluation}
\end{table}
We assessed the degradation score $\mathcal{M}$ across 1.6 million fake and 18,600 original audio samples. Due to the significant imbalance between fake and original samples in our automated evaluation dataset, we rely on AUROC rather than accuracy as our primary metric. This choice ensures a more robust evaluation given the dataset's imbalanced nature.

Table~\ref{tab:automated_evaluation} details our analysis, highlighting the degradation disparity between fake and original samples for all challenges. Fig.~\ref{fig:automated_boxplot} visually demonstrates the separation in machine-scored degradation between original and fake samples.

\vspace{0.35em}\noindent\textbf{Results and Observations.}  Regular speech (no-challenge) exhibited low degradation—11\% for deepfakes and 5\% for originals—indicating RTDFs' proficiency in replicating standard speech patterns.

The introduction of audio challenges led to increased degradation in both deepfakes and original recordings, with deepfakes experiencing more substantial quality deterioration. This effect was particularly pronounced in tasks involving foreign words, static mouth, and cross-talk, resulting in heavy word information loss.

Our examination of 20 prospective challenges revealed 10 to be less effective, with AUROC scores below 70\% (grayed in Table~\ref{tab:automated_evaluation}), leading to their disqualification. These tasks included variations in accent, reading pace, amplitude, pitch, and specific actions like clapping, coughing/whistling, and nose-holding. Deepfake algorithms demonstrated notable proficiency in mimicking tasks related to regular speech patterns, showcasing the sophistication of voice-cloning systems in replicating human vocal characteristics. 

Additional disqualification reasons included: (1) low compliance rate in original recordings for the hold-the-nose challenge (43\%), (2) non-randomizability of the cough/whistle challenge due to lack of spoken content, and (3) false positives in the clapping challenge detector, where deepfakes transformed clap sounds into noisy peaks mimicking clap-like features in the waveform.

\textit{Deepfakes struggled to maintain compliance against top challenges} in scenarios requiring lyrical playback (\#19,\#20), mouth cupping (\#2), whispering (\#3), and high-pitched voice production (\#5). Please refer to Table~\ref{tab:automated_evaluation} and Fig.~\ref{fig:automated_boxplot} (left). These findings indicate a critical vulnerability in current deepfake technologies. Lyrical playback, in particular, introduced measurable vocal distortions, highlighting the effectiveness of these challenges in compromising deepfake audio fidelity.

Our analysis of frequency modulation effects, comparing high-pitch and low-pitch challenges, revealed that high-pitch (\#5) challenges induced more significant degradation (44\%) in deepfake outputs than their low-pitch (\#6) counterparts (25\%). This discrepancy in machine detection efficiency points to \textit{limitations in the training datasets of deepfake models}, such as the VCTK corpus, which lacks diverse high-pitch voice samples. This limitation underscores the critical role of dataset diversity in enhancing deepfake quality.

In addition to FREE-VC~\cite{li2023freevc}, we also evaluated two any-to-many voice converters: StarGANv2-VC~\cite{starganv2vc} and PPG-VC~\cite{ppgvc}. We trained them on our open-sourced dataset to add new identities to their target group. The original implementation of StarGANv2-VC only included 20 speaker identities. In comparison, we trained it on our 100 speaker identities, making them \textit{all seen} during training. The \textit{challenges remain effective} as our automated detectors scored AUC of 88.2\% and 88.3\% against StarGANv2 and PPG, respectively.

In comparison to prior work, D-CAPTCHA~\cite{yasur2023deepfake}, our work identified that six out of their nine tasks were ineffective. As they had evaluated only against StarGANv2-VC~\cite{starganv2vc}, these tasks proved less effective when tested against the more sophisticated FREE-VC model \cite{li2023freevc}, suggesting that our remaining 10 challenges offer higher degradation even against advanced adversaries. 

\vspace{0.35em}\noindent\textbf{Key Takeaway.} We have developed a comprehensive and robust machine evaluation framework, identifying 10 tasks that effectively serve as challenges. These challenges enable machines to achieve an AUROC of 88.7\%, compared to 56.0\% for no-challenge conditions. Our findings highlight the vital role of challenges in identifying deepfake strengths and limitations, particularly when such tasks involve complex auditory challenges beyond simple speech imitation. The analysis reveals two main vulnerabilities of RTDFs: \textit{they are restricted by training data diversity}, and \textit{they do not encode the physiology and physics of human speech apparatus}.

\subsection{RQ2. Vanilla Human Evaluation}

While we have established that challenges enable effective machine evaluations, it is crucial to investigate whether similar results hold for human assessments. System support for human evaluation is essential in scenarios where advanced deepfake detection tools are unavailable, interpretability is desired, or both. This subsection details our vanilla human evaluation and its subsequent extension into two machine-assisted evaluations.

\vspace{0.35em}\noindent\textbf{Creating a Balanced Dataset.} Given the impracticality of manually evaluating 1.6 million audio files, we focused on a hard balanced subset of the data.

Our full dataset comprises 9,900 unique imposter-target speaker pairs. We employed the SpeechBrain speaker recognition framework \cite{speechbrain} to compute voice matching scores ($\in [-1, 1]$) for each pair while maintaining consistent speech content between samples. Of these, 5,643 imposter-target pairs matched at a threshold of 0.25, where a given speech sample is classified as a positive match of the reference. To increase the likelihood of human failure, we further restricted the set by raising the threshold to 0.50. From this subset, we selected 147 high-quality samples per challenge (21 in total), each with a (pMOS) of ($4.50 \pm 0.25$), resulting in a \textit{hard balanced} dataset of 6,174 audio samples for human assessment.

\vspace{0.35em}\noindent\textbf{Setup.} Our human evaluation methodology mirrors the compliance check for original recordings outlined in \S\ref{sec:data_collection}. Therefore, we recruited 63 English-speaking evaluators, categorized into three groups: native monolingual, non-native multilingual, and tie-breaker. We determine the final score for each sample through majority voting across three evaluators.

Evaluators were tasked with assessing each audio sample's compliance with the given task ($\mathcal{C}$) and its overall quality or realism ($\mathcal{R}$) on a 5-point Likert scale, following the ITU P.808 standard for subjective audio quality assessments \cite{itu2018p808} (see Fig.~\ref{fig:compliance_verification}).  

\vspace{0.35em}
We calculate the \textbf{human-scored degradation $\mathcal{H}$} for an audio sample $x$ as follows:
\begin{equation*}
    \mathcal{H}(x) = 1 - \min(\mathcal{C}(x), \mathcal{R}(x) / 5).
\end{equation*}

Here, $\mathcal{C}: x \rightarrow {0,1}$ is 1 if human response was compliant and 0 otherwise. Also $\mathcal{R}(x)$ is a realism rating on a 5-point scale (higher is better). This measure assigns maximum degradation to a non-compliant sample or a score reduction based on human ratings. 

Fig.~\ref{fig:automated_boxplot} (right) illustrates human evaluation results for the top-10 challenges, while Fig.~\ref{fig:human_boxplot} illustrates the distribution of human scores across the 20 challenges and the corresponding AUROC.

\vspace{0.35em}\noindent\textbf{Humans vs Machines.} Humans scored overall 74.7\% AUROC on the 20 challenges, with strong human-human agreement (Kendall's Coefficient $\bar{\tau} = 0.86$). 

Human performance surpassed machines in challenges involving Speaking Loudly, Low Pitch, Cough/Whistle, and Clapping. Conversely, humans were less effective in challenges associated with Static Mouth, Foreign Words, Singing, and Emotion Display, indicating a human sensitivity gap in detecting subtleties within these challenges.
\begin{figure*}[t]
    \centering
    \begin{tabular}{ccc}
        \fbox{\includegraphics[width=0.28\linewidth]{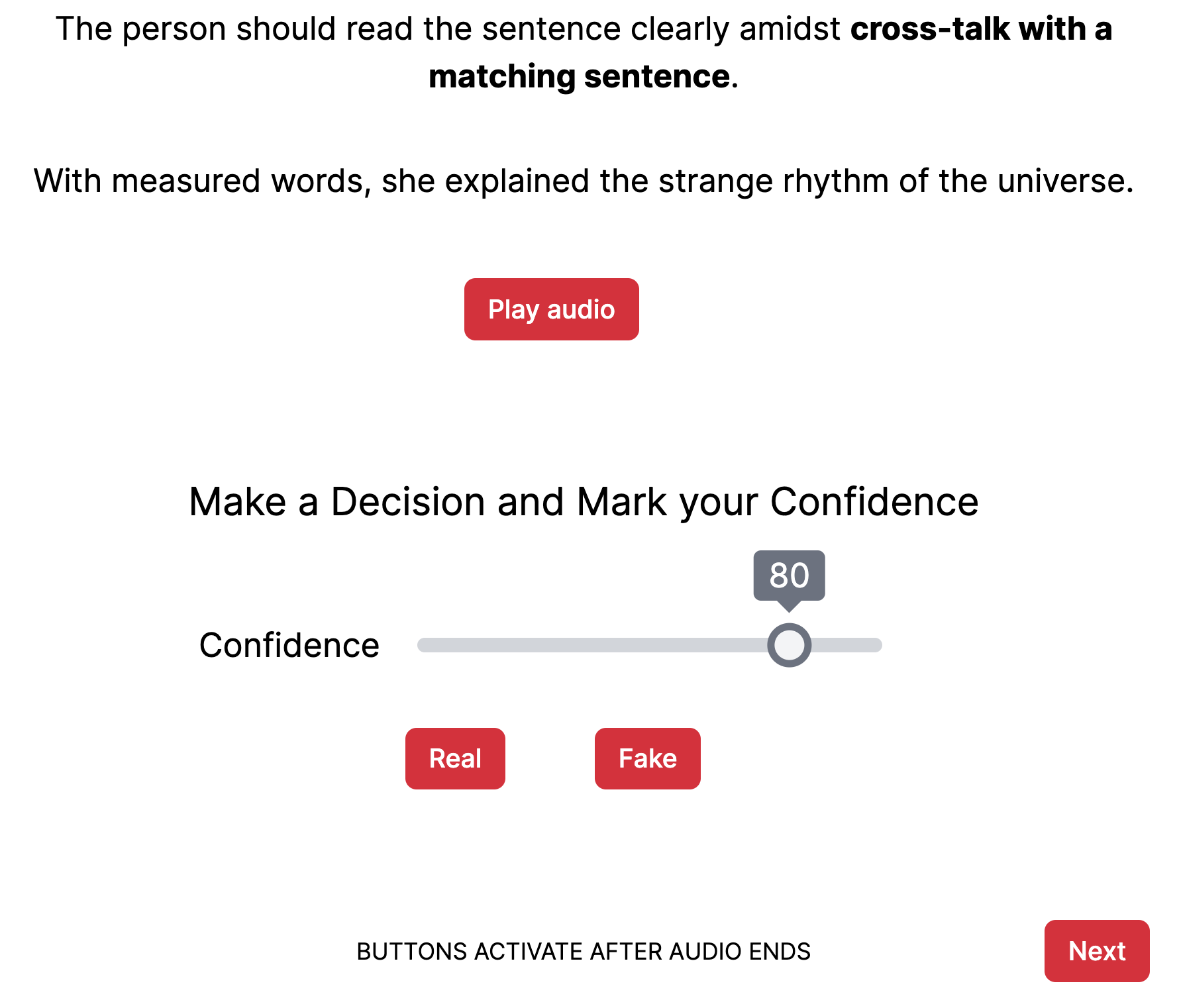}} &
        \fbox{\includegraphics[width=0.28\linewidth]{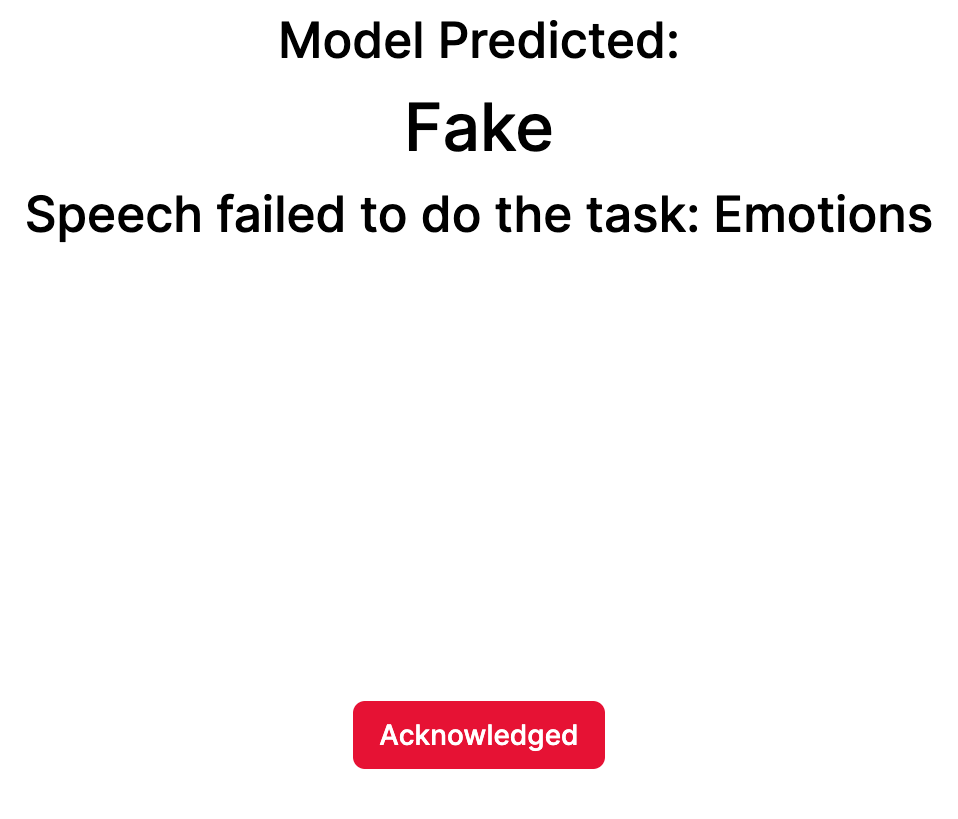}} &
        \fbox{\includegraphics[width=0.28\linewidth]{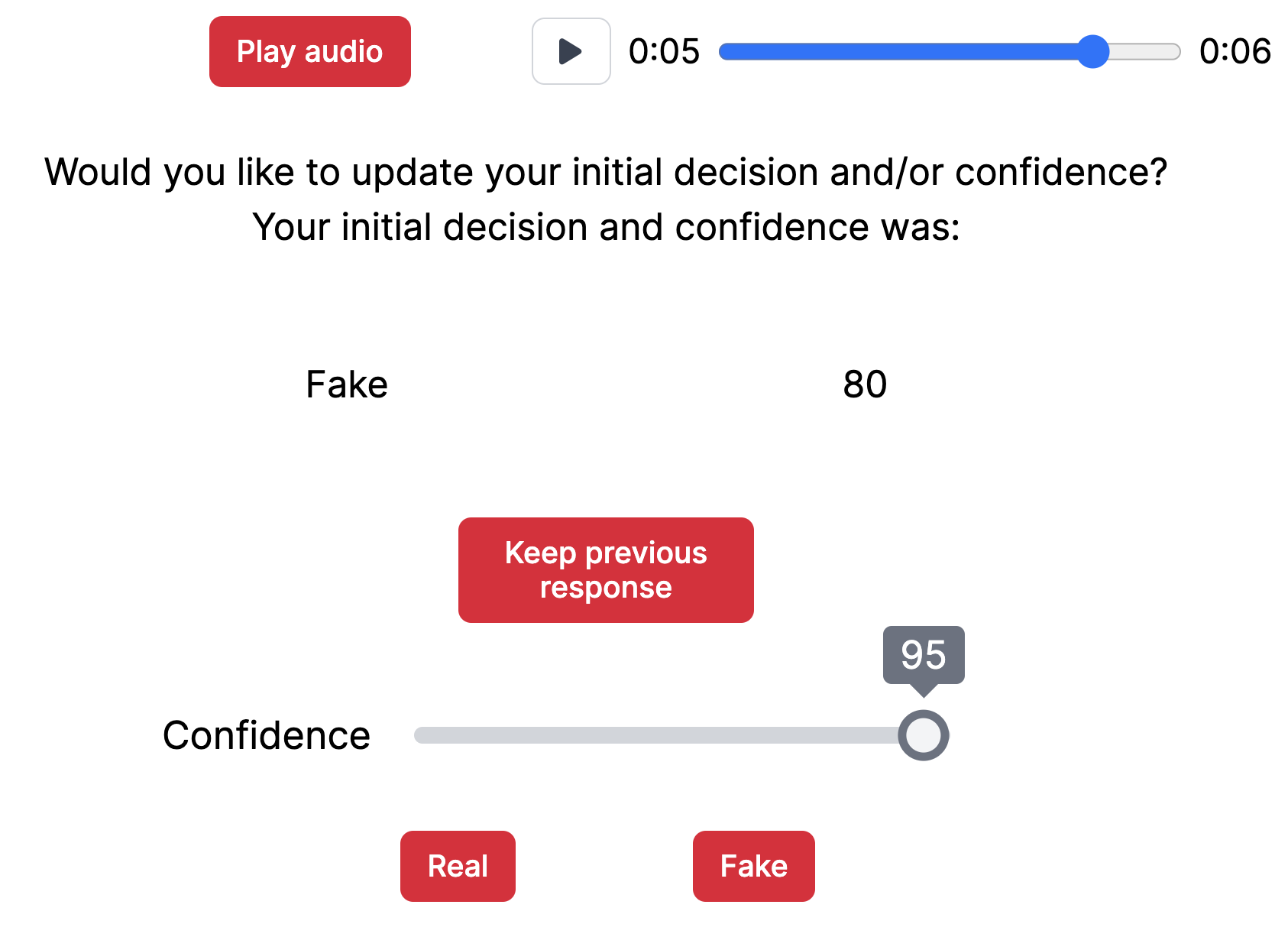}} \\
        (a) Human decision & (b) Machine verdict & (c) Chance to Update decision
    \end{tabular}
    \caption{\small Machine-assisted evaluation process: (a) Humans make an initial decision, (b) Machine verdict is shown, (c) Humans update their decision and confidence.}
    \label{fig:ai_assisted}
\end{figure*}

On the samples corresponding to the top-10 machine-validated challenges in the balanced dataset, human evaluators achieved a mean AUROC of 87.7\% in detecting deepfakes across, comparable to machine evaluations (88.7\%), albeit on a difficult, high-quality dataset. However, their accuracy (not AUROC) was lower at 72.6\% mean accuracy, in comparison to machines' 87.7\%. See breakdown in Table~\ref{tab:human_evaluation} (left). Interestingly, for this set, the Pearson correlation between human and machine AUROC was merely 0.12, implying a poor positive relationship and \textit{revealing the complementary strengths of humans and machines}. 

Also, the machine realism (NISQA) and Humans' MOS ratings on our dataset were uncorrelated or weakly uncorrelated (Pearson Correlation $\rho = -0.05$),\textit{ corroborating our finding that machines and humans are sensitive to different aspects of speech}. See Tab.~\ref{tab:rebuttal_table} for breakdown of coefficients.

\vspace{0.35em}\noindent\textbf{Key Takeaway}. Although the disparity between human and machine performance appears nominal, each exhibits distinct advantages and limitations. This gap becomes \textit{more pronounced when examining the accuracy} on the 10 machine-validated challenges. Here, humans achieved an accuracy of 72.6\% in deepfake detection, compared to 87.7\% by machines. This difference raises the prospect of combining human and machine efforts to enhance overall performance in deepfake detection.

\begin{table}[t!]
    \caption{\small Detection Accuracy of Vanilla, AI-assisted and Collaborative human evaluation, and corresponding percentage boosts for each machine-validated challenge.}
    \centering

    {\small
\begin{tabular}{L{1.68cm}C{1.5cm}|C{0.7cm}C{0.8cm}|C{0.7cm}C{0.9cm}}
\toprule
\textbf{Challenge} & \textbf{Vanilla Acc.} & \multicolumn{2}{c|}{\textbf{Assisted Acc.}} & \multicolumn{2}{c}{\textbf{Collaborative}}\\
\midrule
No Challenge & 52.2 & 66.7 & $\uparrow$27.2\% & \textbf{88.4} & $\uparrow$69.7\% \\
Static Mouth & 58.8 & 68.8 & $\uparrow$16.2\% & \textbf{71.1} & $\uparrow$20.9\% \\
Cup mouth & 84.9 & 91.2 & $\uparrow$7.4\% & \textbf{93.9} & $\uparrow$10.6\% \\
Whisper & 69.6 & 78.1 & $\uparrow$12.6\% & \textbf{81.7} & $\uparrow$17.9\% \\
Speak softly & 61.8 & 63.4 & $\uparrow$2.4\% & \textbf{66.5} & $\uparrow$7.5\% \\
High Pitch & 84.5 & 90.5 & $\uparrow$6.9\% & \textbf{90.6} & $\uparrow$7.1\% \\
Foreign Words & 58.5 & 71.3 & $\uparrow$21.5\% & \textbf{85.4} & $\uparrow$46.3\% \\
Emotions & 68.3 & 74.0 & $\uparrow$7.9\% & \textbf{77.3} & $\uparrow$12.3\% \\
Crosstalk & 80.2 & 85.1 & $\uparrow$6.0\% & \textbf{86.3} & $\uparrow$7.2\% \\
Instr. Playback & 89.2 & 94.6 & $\uparrow$5.8\% & \textbf{95.1} & $\uparrow$6.3\% \\
Lyric Playback & 90.2 & \textbf{90.7} & $\uparrow$0.5\% & 89.1 & $\downarrow$1.2\% \\
\midrule
Average Acc. & 72.6 & 79.4 & $\uparrow$9.3\% & \textbf{84.3} & $\uparrow$16.4\% \\
\bottomrule
\end{tabular}
 }
    \label{tab:human_evaluation}
\end{table}

\subsection{RQ3. AI Assistance \& Human Collaboration}
\label{sec:human_ai_colab}
Given the complementary strengths exhibited by humans and machines, we conducted additional human evaluations to explore the potential of machine assistance in enhancing human capability to detect deepfakes. We focused this investigation on the top 10 challenges (and no-challenge) where machine performance was deemed satisfactory.

For this evaluation, we selected a subset of 92 samples from the original 147 in the hard-balanced dataset. This subsampling resulted in 2,016 deepfake and original audio samples, and allowed us to recruit a larger pool of evaluators.

\vspace{0.35em}\noindent\textbf{Study Design.} We recruited a new, gender-balanced cohort of 100 human evaluators from the Prolific platform. The recruitment process included a short orientation to familiarize participants with the challenge samples they would encounter. We used a preliminary quiz based on original challenge samples to screen participants, with 91 qualifying to serve as an evaluator.

We informed participants about the task and emphasized that the machine's verdict had 87.7\% accuracy (computed by choosing $\mathcal{M}_{chal} = 0.25$ as the threshold for all challenges). Each participant was asked to listen to an audio sample and determine its authenticity ("Real" or "Fake"), as well as their confidence level in this initial decision. After making their initial judgment, we disclosed the machine's verdict to the participant, sometimes accompanied by a brief explanation. We based the explanations on one of three criteria established in \S\ref{sec:machine_evaluation}: failure to perform the specified task, mismatch in spoken text, or presence of vocal distortions. See Fig.~\ref{fig:ai_assisted}. 

After receiving the machine's verdict and rationale, \textit{participants were given the option to revise their decision and confidence level} or stick with their initial assessment. This method generated a total of 8,372 responses.

\vspace{0.35em}\noindent\textbf{Machine-Assisting Humans.} Table~\ref{tab:human_evaluation} (center) details the accuracy of each challenge after machine assistance. We report accuracy rather than AUROC, as the dataset is perfectly balanced. Vanilla decisions achieved a 72.6\% accuracy, while \textit{assisted decisions improved to 79.4\%}, enhancing the vanilla accuracy by 9.3\% overall. Notably, machine assistance increased vanilla human accuracy across all individual challenges. This effect is particularly evident in nuanced challenges such as Static-Mouth, Cupping-Mouth, and Whisper, and less pronounced in scenarios like No-Challenge and Foreign Words (as evaluators were English speakers).

Table~\ref{tab:ai_assisted} presents the initial judgments (I), machine-assists (M), and final decisions (F) across the challenges, breaking down the responses across eight possible scenarios and noting changes in confidence levels. When machine predictions aligned with human decisions, they increased human confidence, regardless of correctness. Conversely, when predictions diverged, they reduced human confidence. Notably, when humans initially erred and machines provided correct guidance, humans changed their decisions 45.4\% of the time (781 / (781 + 939)). However, when humans were correct and machines erred, the latter misled humans 29.8\% of the time (216 / (216 + 510)). This disparity highlights \textit{the beneficial role of machines in correcting human errors} and their relatively lesser impact in misleading humans when they are correct. \textit{We observe no significant difference (42.4\% vs. 43.0\%) in humans accepting the decision} when given the rationale or otherwise.

\begin{figure}[t!]
  \centering
  \includegraphics[width=0.9\columnwidth]{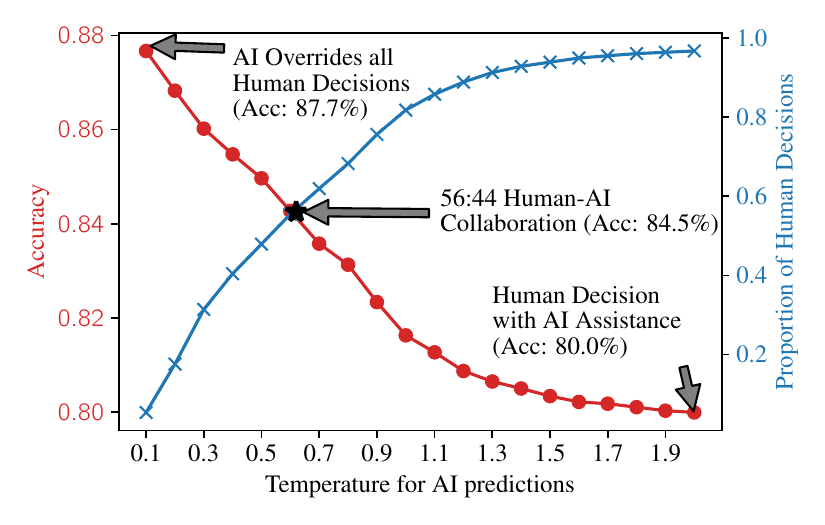}
  \caption{\small Tradeoff between deepfake detection accuracy and human decision retention. AI overrides human decisions when its confidence is higher. Temperature calibrates AI confidence, with lower values increasing AI overrides. The intersection point represents equilibrium between human control and AI automation.}
  \label{fig:human_ai_colab}
\end{figure}

\vspace{0.35em}\noindent\textbf{Human-AI Collaboration.} Our framework implements a confidence based decision routing system that determines when human review is necessary or when machine decision is acceptable. We used a detection threshold $\tau_{base}=0.25$ for classifying samples as deepfakes, accepted across literature~\cite{speechbrain}. For each audio sample, machine confidence is computed as $C_m = |M_{chal}(x) - \tau_{base}|/\tau_{base}$, representing the normalized distance from the decision boundary.

To address the discrepancy between high machine accuracy (87.7\%) and its low confidence, we introduced a temperature parameter $T$ that calibrates confidence scaling: $C_m' = C_m^{1/T}$. This calibration determines which decisions can be automated versus requiring human review. Calibrated samples with high confidence ($C_m' > 0.7$) receive automated decisions, while low-confidence cases ($C_m' \leq 0.7$) are routed to human reviewers with machine predictions, such as `Deepfake-Likely,' displayed as assistive information.

Fig.~\ref{fig:human_ai_colab} illustrates this accuracy-automation trade-off across temperature values. At the optimal operating point ($T=0.7$), humans make 56\% of decisions while machines handle the remaining 44\%, achieving 84.5\% accuracy—a 16.4\% improvement over human-only evaluation (refer Table~\ref{tab:human_evaluation} (right)). This balanced approach ensures human oversight for uncertain cases while leveraging machine precision for high-confidence decisions.

\begin{table}[t!]
    \centering
    \caption{\small Impact of Machine Predictions on Human Decisions. Right column: confidence change. }
    {\footnotesize
    \begin{tabular}{C{3cm}C{0.9cm}C{0.25cm}C{0.3cm}C{0.25cm}C{1.5cm}}
\midrule
\textbf{Scenario} & \textbf{Count}  & \textbf{I} & \textbf{M} & \textbf{F}  & \textbf{$\Delta$Conf.(\%)}\\ \midrule
Correct Agreement & 5362 & \cmark & \cmark & \cmark & $+8.1 \pm 15.8$ \\
Machine Corrected & 781 & \xmark & \cmark & \cmark & $-10.0 \pm 12.6$ \\
Self Corrected & 5 & \xmark & \xmark & \cmark & $-9.9 \pm 20.0$ \\
No Change (Initial Correct) & 510 & \cmark & \xmark & \cmark & $-5.4 \pm 18.9$ \\
Self Misled & 27 & \cmark & \cmark & \xmark & $-6.9 \pm 19.2$ \\
Machine Misled & 216 & \cmark & \xmark & \xmark & $-8.6 \pm 17.1$ \\
No Change (Initial Wrong) & 939 & \xmark & \cmark & \xmark & $-2.2 \pm 17.6$ \\
Incorrect Agreement & 532 & \xmark & \xmark & \xmark & $+8.8 \pm 13.0$ \\ 
\bottomrule
\end{tabular}
    }
    \vspace{0.5em}
    {\footnotesize
    \textbf{Note:} I: Initial decision, M: Machine verdict, F: Final decision. \cmark: correct, \xmark: incorrect.
    }
    \label{tab:ai_assisted}
\end{table}

\vspace{0.35em}\noindent\textbf{Key Takeaway.} Challenges enable human capabilities to detect deepfake audios, similar to machines, even on a significantly challenging dataset. Comparing human detection to machines reveals similarities and differences depending on the challenge type, suggesting a complementary relationship between human intuition and machine precision. When combined, machine tagging aided in \textit{assisting} humans where their performance was lacking. Further collaboration between humans and machines resulted in an even greater boost in deepfake detection performance, while keeping humans in control.

\section{Discussion}
\label{sec:discussion}

\begin{figure}[t!]
  \centering
  \includegraphics[width=0.9\columnwidth]{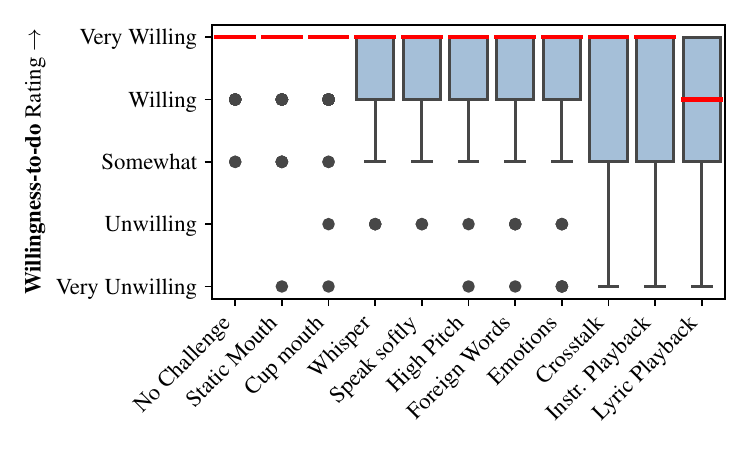}
  \caption{100 users' willingness rated on a 5-point Likert scale}
  \label{fig:usability}
\end{figure}

\vspace{0.35em}\noindent\textbf{Usability.} For a human-centered security system like \textsc{Pitch}, usability is paramount. Fig.~\ref{fig:usability} presents the willingness ratings of the 10 challenges as evaluated by 100 participants who contributed their data. For the caller-side, the median rating for performing all challenges was either Willing' or Very Willing.' As we vet all challenges for effectiveness, organizations can initially deploy our most usable challenges and incorporate others only when necessary.

We acknowledge that our proposed challenge-response framework incorporates manual components and requires callers to actively engage in a task before authentication. However, in terms of execution speed, it typically \textit{takes a caller mere 6 seconds} to verbalize a sentence and complete the challenge.

On the defender's (receiver's) side, there is a desire for swift verification of callers before proceeding with the conversation. To accommodate this, we employed AI as a tagging system to aid the defender in making confident decisions. The design decision to maintain human control acknowledges the fallibility of AI and the nature of telephone communications, where a call, unlike spam emails, cannot be retrieved once disconnected.

\vspace{0.35em}\noindent\textbf{Robustness against Countermeasures.} An adaptive adversary may design countermeasures to counter our approach. 

Firstly, given the public nature of challenges, \textit{adversaries can anticipate a challenge} and prefabricate their responses, utilizing sophisticated editing tools to conceal detectable anomalies. However, \textit{our challenges feature randomizable elements}, such as varying speech content and background noise, introduced at the call's outset, to thwart pre-recordings.

Secondly, as deepfakes advance algorithmically, they may learn to bypass some of our current challenges. Anticipating this, defenders can employ a sequence of existing challenges, akin to current \textsc{Captcha} implementations. Defenders can also incorporate newly discovered challenges into \textsc{Pitch} to stay ahead of technological advancements in deepfake generation. Drawing from our own findings, we assert that there will always exist an audio challenge that deepfakes cannot faithfully generate, as the generative models do not simulate physics or encode human speech physiology.

\vspace{0.35em}
\noindent\textbf{Security-usability tradeoff:} This tradeoff presents two perspectives:
From the user's perspective, we aim to minimize the number of challenges (usability). Conversely, from the defender's perspective, we strive to make confident, accurate decisions (security).

Comparing detection performance (Fig. 2) with usability ratings (Fig. 4), we observe that certain challenges, such as cupping the mouth, whispering, or maintaining a static mouth, best balance this tradeoff. \textit{Even with a single challenge} equivalent to verbalizing a single sentence ($\approx4.5$ seconds), we demonstrate high AUC ($\approx90\%$). Furthermore, if the defender requires greater certainty depending on the sensitivity of the phone interaction, they can request multiple challenges from the user to achieve that level of assurance.

\vspace{0.35em}
\noindent\textbf{Accessibility and Inclusivity Considerations.} \textsc{Pitch}'s vocal challenges could present barriers for speech-impaired individuals as they might be incorrectly flagged as suspicious. We propose an accommodation pathway where users can self-identify speech impairments, triggering automatic human review with alternative authentication methods (pre-registered devices/emails). To prevent exploitation of this pathway, machine evaluation should still run in parallel, providing reviewers with supplementary data for informed decision-making.

%


Also, though our evaluation focused on English speakers, \textsc{Pitch}'s taxonomy is fundamentally language-agnostic and extends readily to multilingual implementations. Notably, accent diversity among non-native speakers strengthens the system's effectiveness, as acoustic variations in accented speech create additional complexity for voice conversion systems trained on homogeneous data.

Hence, \textsc{Pitch} balances accessibility needs with security requirements, while being inclusive and maintaining verification integrity.

\vspace{0.35em}
\noindent\textbf{Limitations and Future Work.} 

\begin{itemize}[nosep,leftmargin=*]
\item \textit{Comprehensive challenge evaluation}: Our study did not assess all sub-categories in our challenge taxonomy, particularly those that might prove more relevant in a multi-lingual context.
    \item \textit{Testing against COTS deepfake generators:} \textsc{Pitch} has not been tested against commercial off-the-shelf deepfake generators.
    \item \textit{Advanced ML models:} Utilize more sophisticated and recent foundation models to enhance cross-lingual machine evaluation.
\end{itemize}

\section{Conclusion}
\label{sec:conclusion}
This study presents a novel challenge-response approach, \textsc{Pitch}, to detect interactive deepfake audio calls in real-world authentication scenarios. We developed a comprehensive taxonomy of audio challenges, identifying 10 robust instances from 20 candidates, and created a diverse open-source dataset of 18,600 original and 1.6 million deepfake samples. These challenges effectively degraded deepfake quality, improving machine detection capabilities from 56.0\% to 88.7\% AUROC and enabling humans to discern deepfakes with 72.6\% accuracy, even against a challenging dataset baseline.

By fostering a synergistic relationship between humans and machines, we achieved a balanced framework that capitalizes on human oversight and machine efficiency. Our collaborative approach reached 84.5\% detection accuracy while maintaining human decision control in 56\% of cases—a significant improvement over individual performance. This methodology provides a practical solution that requires no prior enrollment or reference samples, laying the foundation for robust defensive measures against increasingly sophisticated voice-based social engineering attacks across government, healthcare and finance sectors.

\bibliographystyle{ACM-Reference-Format}
\bibliography{newbib}


\begin{thebibliography}{49}


\ifx \showCODEN    \undefined \def \showCODEN     #1{\unskip}     \fi
\ifx \showISBNx    \undefined \def \showISBNx     #1{\unskip}     \fi
\ifx \showISBNxiii \undefined \def \showISBNxiii  #1{\unskip}     \fi
\ifx \showISSN     \undefined \def \showISSN      #1{\unskip}     \fi
\ifx \showLCCN     \undefined \def \showLCCN      #1{\unskip}     \fi
\ifx \shownote     \undefined \def \shownote      #1{#1}          \fi
\ifx \showarticletitle \undefined \def \showarticletitle #1{#1}   \fi
\ifx \showURL      \undefined \def \showURL       {\relax}        \fi
\providecommand\bibfield[2]{#2}
\providecommand\bibinfo[2]{#2}
\providecommand\natexlab[1]{#1}
\providecommand\showeprint[2][]{arXiv:#2}

\bibitem[Altalahin et~al\mbox{.}(2023)]%
        {altalahin2023unmasking}
\bibfield{author}{\bibinfo{person}{Islam Altalahin}, \bibinfo{person}{Shadi AlZu'bi}, \bibinfo{person}{Assal Alqudah}, {and} \bibinfo{person}{Ala Mughaid}.} \bibinfo{year}{2023}\natexlab{}.
\newblock \showarticletitle{Unmasking the Truth: A Deep Learning Approach to Detecting Deepfake Audio Through MFCC Features}. In \bibinfo{booktitle}{\emph{2023 International Conference on Information Technology (ICIT)}}. IEEE, \bibinfo{pages}{511--518}.
\newblock


\bibitem[APNews(2023)]%
        {nycmayordeepfakes}
\bibfield{author}{\bibinfo{person}{APNews}.} \bibinfo{year}{2023}\natexlab{}.
\newblock \bibinfo{title}{{Can New York's mayor speak Mandarin? No, but with AI he's making robocalls in different languages}}.
\newblock
\urldef\tempurl%
\url{https://apnews.com/article/nyc-mayor-ai-robocalls-foreign-languages-30517885466994e5f1f54745c08691e0}
\showURL{%
\tempurl}
\newblock
\shownote{Accessed: 23-Nov-2023}.


\bibitem[Baevski et~al\mbox{.}(2020)]%
        {wav2vec2020}
\bibfield{author}{\bibinfo{person}{Alexei Baevski}, \bibinfo{person}{Yuhao Zhou}, \bibinfo{person}{Abdelrahman Mohamed}, {and} \bibinfo{person}{Michael Auli}.} \bibinfo{year}{2020}\natexlab{}.
\newblock \showarticletitle{wav2vec 2.0: A framework for self-supervised learning of speech representations}.
\newblock \bibinfo{journal}{\emph{Advances in neural information processing systems}}  \bibinfo{volume}{33} (\bibinfo{year}{2020}), \bibinfo{pages}{12449--12460}.
\newblock


\bibitem[Bain et~al\mbox{.}(2023)]%
        {bain2022whisperx}
\bibfield{author}{\bibinfo{person}{Max Bain}, \bibinfo{person}{Jaesung Huh}, \bibinfo{person}{Tengda Han}, {and} \bibinfo{person}{Andrew Zisserman}.} \bibinfo{year}{2023}\natexlab{}.
\newblock \showarticletitle{WhisperX: Time-Accurate Speech Transcription of Long-Form Audio}. In \bibinfo{booktitle}{\emph{Interspeech 2023}}. \bibinfo{pages}{4489--4493}.
\newblock
\showISSN{2958-1796}
\href{https://doi.org/10.21437/Interspeech.2023-78}{doi:\nolinkurl{10.21437/Interspeech.2023-78}}


\bibitem[Betker(2023)]%
        {tortoise_tts}
\bibfield{author}{\bibinfo{person}{James Betker}.} \bibinfo{year}{2023}\natexlab{}.
\newblock \showarticletitle{Better speech synthesis through scaling}.
\newblock \bibinfo{journal}{\emph{arXiv preprint arXiv:2305.07243}} (\bibinfo{year}{2023}).
\newblock


\bibitem[Blue et~al\mbox{.}(2022)]%
        {blue2022you}
\bibfield{author}{\bibinfo{person}{Logan Blue}, \bibinfo{person}{Kevin Warren}, \bibinfo{person}{Hadi Abdullah}, \bibinfo{person}{Cassidy Gibson}, \bibinfo{person}{Luis Vargas}, \bibinfo{person}{Jessica O{\textquoteright}Dell}, \bibinfo{person}{Kevin Butler}, {and} \bibinfo{person}{Patrick Traynor}.} \bibinfo{year}{2022}\natexlab{}.
\newblock \showarticletitle{Who Are You (I Really Wanna Know)? Detecting Audio {DeepFakes} Through Vocal Tract Reconstruction}. In \bibinfo{booktitle}{\emph{31st USENIX Security Symposium (USENIX Security 22)}}. \bibinfo{publisher}{USENIX Association}, \bibinfo{address}{Boston, MA}, \bibinfo{pages}{2691--2708}.
\newblock
\showISBNx{978-1-939133-31-1}
\urldef\tempurl%
\url{https://www.usenix.org/conference/usenixsecurity22/presentation/blue}
\showURL{%
\tempurl}


\bibitem[Chowdhury et~al\mbox{.}(2022)]%
        {singing2022icassp}
\bibfield{author}{\bibinfo{person}{Anurag Chowdhury}, \bibinfo{person}{Austin Cozzo}, {and} \bibinfo{person}{Arun Ross}.} \bibinfo{year}{2022}\natexlab{}.
\newblock \showarticletitle{Domain Adaptation for Speaker Recognition in Singing and Spoken Voice}. In \bibinfo{booktitle}{\emph{ICASSP 2022 - 2022 IEEE International Conference on Acoustics, Speech and Signal Processing (ICASSP)}}. \bibinfo{pages}{7192--7196}.
\newblock
\href{https://doi.org/10.1109/ICASSP43922.2022.9746111}{doi:\nolinkurl{10.1109/ICASSP43922.2022.9746111}}


\bibitem[{CNN}(2023)]%
        {kidnappergirl}
\bibfield{author}{\bibinfo{person}{{CNN}}.} \bibinfo{year}{2023}\natexlab{}.
\newblock \bibinfo{title}{{'Mom, these bad men have me': She believes scammers cloned her daughter's voice in a fake kidnapping}}.
\newblock
\urldef\tempurl%
\url{https://www.cnn.com/2023/04/29/us/ai-scam-calls-kidnapping-cec/index.html}
\showURL{%
\tempurl}
\newblock
\shownote{Accessed: 23-Nov-2023}.


\bibitem[{CNN}(2024)]%
        {cnn_deepfake_scam_2024}
\bibfield{author}{\bibinfo{person}{{CNN}}.} \bibinfo{year}{2024}\natexlab{}.
\newblock \bibinfo{title}{Hong Kong finance worker loses \$25 million in deepfake video conference scam}.
\newblock
\urldef\tempurl%
\url{https://www.cnn.com/2024/02/04/asia/deepfake-cfo-scam-hong-kong-intl-hnk/index.html}
\showURL{%
\tempurl}
\newblock
\shownote{Accessed: 2024-02-15}.


\bibitem[{CSO Online}(2023)]%
        {csoonline_generative_ai_kyc}
\bibfield{author}{\bibinfo{person}{{CSO Online}}.} \bibinfo{year}{2023}\natexlab{}.
\newblock \bibinfo{title}{Will Generative AI Kill KYC Authentication?}
\newblock
\urldef\tempurl%
\url{https://www.csoonline.com/article/1307021/will-generative-ai-kill-kyc-authentication.html}
\showURL{%
\tempurl}
\newblock
\shownote{Accessed: 2024-02-15}.


\bibitem[Deng et~al\mbox{.}(2024)]%
        {10446798}
\bibfield{author}{\bibinfo{person}{Junlong Deng}, \bibinfo{person}{Yanzhen Ren}, \bibinfo{person}{Tong Zhang}, \bibinfo{person}{Hongcheng Zhu}, {and} \bibinfo{person}{Zongkun Sun}.} \bibinfo{year}{2024}\natexlab{}.
\newblock \showarticletitle{VFD-Net: Vocoder Fingerprints Detection for Fake Audio}. In \bibinfo{booktitle}{\emph{ICASSP 2024 - 2024 IEEE International Conference on Acoustics, Speech and Signal Processing (ICASSP)}}. \bibinfo{pages}{12151--12155}.
\newblock
\href{https://doi.org/10.1109/ICASSP48485.2024.10446798}{doi:\nolinkurl{10.1109/ICASSP48485.2024.10446798}}


\bibitem[{Entrust}(2025)]%
        {entrust2025}
\bibfield{author}{\bibinfo{person}{{Entrust}}.} \bibinfo{year}{2025}\natexlab{}.
\newblock \bibinfo{booktitle}{\emph{2025 Identity Fraud Report}}.
\newblock \bibinfo{type}{Technical Report}. \bibinfo{institution}{Entrust}.
\newblock
\urldef\tempurl%
\url{https://www.entrust.com/sites/default/files/documentation/reports/2025-identity-fraud-report.pdf}
\showURL{%
\tempurl}


\bibitem[Fanelle et~al\mbox{.}(2020)]%
        {fanelle2020blind}
\bibfield{author}{\bibinfo{person}{Valerie Fanelle}, \bibinfo{person}{Sepideh Karimi}, \bibinfo{person}{Aditi Shah}, \bibinfo{person}{Bharath Subramanian}, {and} \bibinfo{person}{Sauvik Das}.} \bibinfo{year}{2020}\natexlab{}.
\newblock \showarticletitle{Blind and human: Exploring more usable audio $\{$CAPTCHA$\}$ designs}. In \bibinfo{booktitle}{\emph{Sixteenth Symposium on Usable Privacy and Security (SOUPS 2020)}}. \bibinfo{pages}{111--125}.
\newblock


\bibitem[ITU(2018)]%
        {itu2018p808}
\bibfield{author}{\bibinfo{person}{Rec ITU}.} \bibinfo{year}{2018}\natexlab{}.
\newblock \showarticletitle{P. 808: Subjevtive Evaluation of Speech Quality With a Crowdsoucing Approach}.
\newblock \bibinfo{journal}{\emph{International Telecommunication Standardization Sector (ITU-T)}} (\bibinfo{year}{2018}).
\newblock


\bibitem[Kawa et~al\mbox{.}(2022)]%
        {kawa2022specrnet}
\bibfield{author}{\bibinfo{person}{Piotr Kawa}, \bibinfo{person}{Marcin Plata}, {and} \bibinfo{person}{Piotr Syga}.} \bibinfo{year}{2022}\natexlab{}.
\newblock \showarticletitle{Specrnet: Towards faster and more accessible audio deepfake detection}. In \bibinfo{booktitle}{\emph{2022 IEEE International Conference on Trust, Security and Privacy in Computing and Communications (TrustCom)}}. IEEE, \bibinfo{pages}{792--799}.
\newblock


\bibitem[Korshunov et~al\mbox{.}(2023)]%
        {korshunov2023vulnerability}
\bibfield{author}{\bibinfo{person}{Pavel Korshunov}, \bibinfo{person}{Haolin Chen}, \bibinfo{person}{Philip~N Garner}, {and} \bibinfo{person}{Sebastien Marcel}.} \bibinfo{year}{2023}\natexlab{}.
\newblock \showarticletitle{Vulnerability of Automatic Identity Recognition to Audio-Visual Deepfakes}.
\newblock  (\bibinfo{year}{2023}).
\newblock


\bibitem[{KYC AML Guide}(2023)]%
        {kycaml_biometrics_voice_recognition}
\bibfield{author}{\bibinfo{person}{{KYC AML Guide}}.} \bibinfo{year}{2023}\natexlab{}.
\newblock \bibinfo{title}{How Does Biometrics Voice Recognition Work?}
\newblock
\urldef\tempurl%
\url{https://kycaml.guide/blog/how-does-biometrics-voice-recognition-work/}
\showURL{%
\tempurl}
\newblock
\shownote{Accessed: 2024-02-15}.


\bibitem[Li et~al\mbox{.}(2023b)]%
        {li2023freevc}
\bibfield{author}{\bibinfo{person}{Jingyi Li}, \bibinfo{person}{Weiping Tu}, {and} \bibinfo{person}{Li Xiao}.} \bibinfo{year}{2023}\natexlab{b}.
\newblock \showarticletitle{Freevc: Towards High-Quality Text-Free One-Shot Voice Conversion}. In \bibinfo{booktitle}{\emph{ICASSP 2023-2023 IEEE International Conference on Acoustics, Speech and Signal Processing (ICASSP)}}. IEEE, \bibinfo{pages}{1--5}.
\newblock


\bibitem[Li et~al\mbox{.}(2024)]%
        {safeear}
\bibfield{author}{\bibinfo{person}{Xinfeng Li}, \bibinfo{person}{Kai Li}, \bibinfo{person}{Yifan Zheng}, \bibinfo{person}{Chen Yan}, \bibinfo{person}{Xiaoyu Ji}, {and} \bibinfo{person}{Wenyuan Xu}.} \bibinfo{year}{2024}\natexlab{}.
\newblock \showarticletitle{SafeEar: Content Privacy-Preserving Audio Deepfake Detection}. In \bibinfo{booktitle}{\emph{Proceedings of the 2024 ACM SIGSAC Conference on Computer and Communications Security}} \emph{(\bibinfo{series}{CCS '24})}. \bibinfo{publisher}{Association for Computing Machinery}, \bibinfo{address}{New York, NY, USA}.
\newblock


\bibitem[Li et~al\mbox{.}(2023a)]%
        {Li2023StyleTTS2T}
\bibfield{author}{\bibinfo{person}{Yinghao~Aaron Li}, \bibinfo{person}{Cong Han}, \bibinfo{person}{Vinay~S. Raghavan}, \bibinfo{person}{Gavin Mischler}, {and} \bibinfo{person}{Nima Mesgarani}.} \bibinfo{year}{2023}\natexlab{a}.
\newblock \showarticletitle{StyleTTS 2: Towards Human-Level Text-to-Speech through Style Diffusion and Adversarial Training with Large Speech Language Models}. In \bibinfo{booktitle}{\emph{Advances in Neural Information Processing Systems}}.
\newblock
\urldef\tempurl%
\url{https://api.semanticscholar.org/CorpusID:259145293}
\showURL{%
\tempurl}


\bibitem[Li et~al\mbox{.}(2021)]%
        {starganv2vc}
\bibfield{author}{\bibinfo{person}{Yinghao~Aaron Li}, \bibinfo{person}{Ali Zare}, {and} \bibinfo{person}{Nima Mesgarani}.} \bibinfo{year}{2021}\natexlab{}.
\newblock \showarticletitle{{StarGANv2-VC: A Diverse, Unsupervised, Non-Parallel Framework for Natural-Sounding Voice Conversion}}. In \bibinfo{booktitle}{\emph{Proc. Interspeech 2021}}. \bibinfo{pages}{1349--1353}.
\newblock
\href{https://doi.org/10.21437/Interspeech.2021-319}{doi:\nolinkurl{10.21437/Interspeech.2021-319}}


\bibitem[Liu et~al\mbox{.}(2021)]%
        {ppgvc}
\bibfield{author}{\bibinfo{person}{Songxiang Liu}, \bibinfo{person}{Yuewen Cao}, \bibinfo{person}{Disong Wang}, \bibinfo{person}{Xixin Wu}, \bibinfo{person}{Xunying Liu}, {and} \bibinfo{person}{Helen Meng}.} \bibinfo{year}{2021}\natexlab{}.
\newblock \showarticletitle{Any-to-Many Voice Conversion With Location-Relative Sequence-to-Sequence Modeling}.
\newblock \bibinfo{journal}{\emph{IEEE/ACM Transactions on Audio, Speech, and Language Processing}}  \bibinfo{volume}{29} (\bibinfo{year}{2021}), \bibinfo{pages}{1717--1728}.
\newblock
\href{https://doi.org/10.1109/TASLP.2021.3076867}{doi:\nolinkurl{10.1109/TASLP.2021.3076867}}


\bibitem[Liu et~al\mbox{.}(2024)]%
        {liu2024generalizing}
\bibfield{author}{\bibinfo{person}{Xuechen Liu}, \bibinfo{person}{Md Sahidullah}, \bibinfo{person}{Kong~Aik Lee}, {and} \bibinfo{person}{Tomi Kinnunen}.} \bibinfo{year}{2024}\natexlab{}.
\newblock \showarticletitle{Generalizing speaker verification for spoof awareness in the embedding space}.
\newblock \bibinfo{journal}{\emph{IEEE/ACM Transactions on Audio, Speech, and Language Processing}}  \bibinfo{volume}{32} (\bibinfo{year}{2024}), \bibinfo{pages}{1261--1273}.
\newblock


\bibitem[Liu et~al\mbox{.}(2018)]%
        {liu2018learning}
\bibfield{author}{\bibinfo{person}{Yaojie Liu}, \bibinfo{person}{Amin Jourabloo}, {and} \bibinfo{person}{Xiaoming Liu}.} \bibinfo{year}{2018}\natexlab{}.
\newblock \showarticletitle{Learning deep models for face anti-spoofing: Binary or auxiliary supervision}. In \bibinfo{booktitle}{\emph{Proceedings of the IEEE conference on computer vision and pattern recognition}}. \bibinfo{pages}{389--398}.
\newblock


\bibitem[McAfee({[n.\,d.]})]%
        {mcafee}
\bibfield{author}{\bibinfo{person}{McAfee}.} \bibinfo{year}{[n.\,d.]}\natexlab{}.
\newblock \bibinfo{title}{{Beware the Artificial Impostor}}.
\newblock
\urldef\tempurl%
\url{https://www.mcafee.com/content/dam/consumer/en-us/resources/cybersecurity/artificial-intelligence/rp-beware-the-artificial-impostor-report.pdf}
\showURL{%
\tempurl}
\newblock
\shownote{Accessed: 23-Nov-2023}.


\bibitem[McGurk and MacDonald(1976)]%
        {mcgurk1976hearing}
\bibfield{author}{\bibinfo{person}{Harry McGurk} {and} \bibinfo{person}{John MacDonald}.} \bibinfo{year}{1976}\natexlab{}.
\newblock \showarticletitle{Hearing lips and seeing voices}.
\newblock \bibinfo{journal}{\emph{Nature}} \bibinfo{volume}{264}, \bibinfo{number}{5588} (\bibinfo{year}{1976}), \bibinfo{pages}{746--748}.
\newblock


\bibitem[Mittag et~al\mbox{.}(2021)]%
        {NISQA}
\bibfield{author}{\bibinfo{person}{Gabriel Mittag}, \bibinfo{person}{Babak Naderi}, \bibinfo{person}{Assmaa Chehadi}, {and} \bibinfo{person}{Sebastian Möller}.} \bibinfo{year}{2021}\natexlab{}.
\newblock \showarticletitle{NISQA: A Deep CNN-Self-Attention Model for Multidimensional Speech Quality Prediction with Crowdsourced Datasets}. In \bibinfo{booktitle}{\emph{Interspeech 2021}}. \bibinfo{publisher}{ISCA}.
\newblock
\href{https://doi.org/10.21437/interspeech.2021-299}{doi:\nolinkurl{10.21437/interspeech.2021-299}}


\bibitem[Mittal et~al\mbox{.}(2024)]%
        {mittal2022gotcha}
\bibfield{author}{\bibinfo{person}{Govind Mittal}, \bibinfo{person}{Chinmay Hegde}, {and} \bibinfo{person}{Nasir Memon}.} \bibinfo{year}{2024}\natexlab{}.
\newblock \showarticletitle{GOTCHA: Real-Time Video Deepfake Detection via Challenge-Response}. In \bibinfo{booktitle}{\emph{2024 IEEE European Symposium on Security and Privacy (EuroS\&P)}}. IEEE.
\newblock


\bibitem[Morris et~al\mbox{.}(2004)]%
        {word-info-lost}
\bibfield{author}{\bibinfo{person}{Andrew Morris}, \bibinfo{person}{Viktoria Maier}, {and} \bibinfo{person}{Phil Green}.} \bibinfo{year}{2004}\natexlab{}.
\newblock \showarticletitle{From WER and RIL to MER and WIL: improved evaluation measures for connected speech recognition}.
\newblock
\href{https://doi.org/10.21437/Interspeech.2004-668}{doi:\nolinkurl{10.21437/Interspeech.2004-668}}


\bibitem[M{\"u}ller et~al\mbox{.}(2022)]%
        {muller2022human}
\bibfield{author}{\bibinfo{person}{Nicolas~M M{\"u}ller}, \bibinfo{person}{Karla Pizzi}, {and} \bibinfo{person}{Jennifer Williams}.} \bibinfo{year}{2022}\natexlab{}.
\newblock \showarticletitle{Human perception of audio deepfakes}. In \bibinfo{booktitle}{\emph{Proceedings of the 1st International Workshop on Deepfake Detection for Audio Multimedia}}. \bibinfo{pages}{85--91}.
\newblock


\bibitem[Nautsch et~al\mbox{.}(2021)]%
        {nautsch2021asvspoof}
\bibfield{author}{\bibinfo{person}{Andreas Nautsch}, \bibinfo{person}{Xin Wang}, \bibinfo{person}{Nicholas Evans}, \bibinfo{person}{Tomi~H Kinnunen}, \bibinfo{person}{Ville Vestman}, \bibinfo{person}{Massimiliano Todisco}, \bibinfo{person}{H{\'e}ctor Delgado}, \bibinfo{person}{Md Sahidullah}, \bibinfo{person}{Junichi Yamagishi}, {and} \bibinfo{person}{Kong~Aik Lee}.} \bibinfo{year}{2021}\natexlab{}.
\newblock \showarticletitle{ASVspoof 2019: spoofing countermeasures for the detection of synthesized, converted and replayed speech}.
\newblock \bibinfo{journal}{\emph{IEEE Transactions on Biometrics, Behavior, and Identity Science}} \bibinfo{volume}{3}, \bibinfo{number}{2} (\bibinfo{year}{2021}), \bibinfo{pages}{252--265}.
\newblock


\bibitem[{NBC News}(2024)]%
        {nbcnews_biden_robocall_2024}
\bibfield{author}{\bibinfo{person}{{NBC News}}.} \bibinfo{year}{2024}\natexlab{}.
\newblock \bibinfo{title}{Fake Joe Biden Robocall Tells New Hampshire Democrats Not to Vote Tuesday}.
\newblock
\urldef\tempurl%
\url{https://www.nbcnews.com/politics/2024-election/fake-joe-biden-robocall-tells-new-hampshire-democrats-not-vote-tuesday-rcna134984}
\showURL{%
\tempurl}
\newblock
\shownote{Accessed: 2024-02-15}.


\bibitem[Nguyen-Le et~al\mbox{.}(2024)]%
        {dcaptchaplusplus}
\bibfield{author}{\bibinfo{person}{Hong-Hanh Nguyen-Le}, \bibinfo{person}{Van-Tuan Tran}, \bibinfo{person}{Dinh-Thuc Nguyen}, {and} \bibinfo{person}{Nhien-An Le-Khac}.} \bibinfo{year}{2024}\natexlab{}.
\newblock \showarticletitle{D-CAPTCHA++: A Study of Resilience of Deepfake CAPTCHA under Transferable Imperceptible Adversarial Attack}. In \bibinfo{booktitle}{\emph{2024 International Joint Conference on Neural Networks (IJCNN)}}. \bibinfo{pages}{1--8}.
\newblock
\href{https://doi.org/10.1109/IJCNN60899.2024.10650401}{doi:\nolinkurl{10.1109/IJCNN60899.2024.10650401}}


\bibitem[{NYTimes}(2023)]%
        {nytimes_voice_deepfakes}
\bibfield{author}{\bibinfo{person}{{NYTimes}}.} \bibinfo{year}{2023}\natexlab{}.
\newblock \showarticletitle{Voice Deepfakes Are Coming for Your Bank Balance}.
\newblock \bibinfo{journal}{\emph{The New York Times}} (\bibinfo{date}{30 Aug} \bibinfo{year}{2023}).
\newblock
\urldef\tempurl%
\url{https://www.nytimes.com/2023/08/30/business/voice-deepfakes-bank-scams.html}
\showURL{%
\tempurl}
\newblock
\shownote{Accessed: 2024-02-15}.


\bibitem[Ravanelli et~al\mbox{.}(2021)]%
        {speechbrain}
\bibfield{author}{\bibinfo{person}{Mirco Ravanelli}, \bibinfo{person}{Titouan Parcollet}, \bibinfo{person}{Peter Plantinga}, \bibinfo{person}{Aku Rouhe}, \bibinfo{person}{Samuele Cornell}, \bibinfo{person}{Loren Lugosch}, \bibinfo{person}{Cem Subakan}, \bibinfo{person}{Nauman Dawalatabad}, \bibinfo{person}{Abdelwahab Heba}, \bibinfo{person}{Jianyuan Zhong}, \bibinfo{person}{Ju-Chieh Chou}, \bibinfo{person}{Sung-Lin Yeh}, \bibinfo{person}{Szu-Wei Fu}, \bibinfo{person}{Chien-Feng Liao}, \bibinfo{person}{Elena Rastorgueva}, \bibinfo{person}{François Grondin}, \bibinfo{person}{William Aris}, \bibinfo{person}{Hwidong Na}, \bibinfo{person}{Yan Gao}, \bibinfo{person}{Renato De~Mori}, {and} \bibinfo{person}{Yoshua Bengio}.} \bibinfo{year}{2021}\natexlab{}.
\newblock \bibinfo{title}{{SpeechBrain}: A General-Purpose Speech Toolkit}.
\newblock
\showeprint[arxiv]{2106.04624}~[eess.AS]
\newblock
\shownote{arXiv:2106.04624}.


\bibitem[Rosenberg and Ramabhadran(2017)]%
        {rosenberg2017bias}
\bibfield{author}{\bibinfo{person}{Andrew Rosenberg} {and} \bibinfo{person}{Bhuvana Ramabhadran}.} \bibinfo{year}{2017}\natexlab{}.
\newblock \showarticletitle{Bias and Statistical Significance in Evaluating Speech Synthesis with Mean Opinion Scores}. In \bibinfo{booktitle}{\emph{Interspeech}}. \bibinfo{pages}{3976--3980}.
\newblock


\bibitem[{suno-ai}(2023)]%
        {bark_suno_ai}
\bibfield{author}{\bibinfo{person}{{suno-ai}}.} \bibinfo{year}{2023}\natexlab{}.
\newblock \bibinfo{title}{{Bark: Text-Prompted Generative Audio Model}}.
\newblock
\urldef\tempurl%
\url{https://github.com/suno-ai/bark}
\showURL{%
\tempurl}
\newblock
\shownote{Accessed: 2024-02-15}.


\bibitem[Tak et~al\mbox{.}(2021)]%
        {rawnet2}
\bibfield{author}{\bibinfo{person}{Hemlata Tak}, \bibinfo{person}{Jose Patino}, \bibinfo{person}{Massimiliano Todisco}, \bibinfo{person}{Andreas Nautsch}, \bibinfo{person}{Nicholas Evans}, {and} \bibinfo{person}{Anthony Larcher}.} \bibinfo{year}{2021}\natexlab{}.
\newblock \showarticletitle{End-to-end anti-spoofing with rawnet2}. In \bibinfo{booktitle}{\emph{ICASSP 2021-2021 IEEE International Conference on Acoustics, Speech and Signal Processing (ICASSP)}}. IEEE, \bibinfo{pages}{6369--6373}.
\newblock


\bibitem[Todisco et~al\mbox{.}(2019)]%
        {todisco2019asvspoof}
\bibfield{author}{\bibinfo{person}{Massimiliano Todisco}, \bibinfo{person}{Xin Wang}, \bibinfo{person}{Ville Vestman}, \bibinfo{person}{Md Sahidullah}, \bibinfo{person}{H{\'e}ctor Delgado}, \bibinfo{person}{Andreas Nautsch}, \bibinfo{person}{Junichi Yamagishi}, \bibinfo{person}{Nicholas Evans}, \bibinfo{person}{Tomi Kinnunen}, {and} \bibinfo{person}{Kong~Aik Lee}.} \bibinfo{year}{2019}\natexlab{}.
\newblock \showarticletitle{ASVspoof 2019: Future horizons in spoofed and fake audio detection}.
\newblock \bibinfo{journal}{\emph{arXiv preprint arXiv:1904.05441}} (\bibinfo{year}{2019}).
\newblock


\bibitem[Truecaller(2024)]%
        {TruecallerSpamScam2024}
\bibfield{author}{\bibinfo{person}{Truecaller}.} \bibinfo{year}{2024}\natexlab{}.
\newblock \bibinfo{booktitle}{\emph{America Under Attack: The Shifting Landscape of Spam and Scam Calls in America}}.
\newblock \bibinfo{type}{{T}echnical {R}eport}. \bibinfo{institution}{Public Interest Network}.
\newblock
\urldef\tempurl%
\url{https://publicinterestnetwork.org/wp-content/uploads/2024/10/US-SpamScam-Report_2024_0307.pdf}
\showURL{%
\tempurl}
\newblock
\shownote{Accessed on February 27, 2025}.


\bibitem[{Wall Street Journal}(2019)]%
        {energyscam}
\bibfield{author}{\bibinfo{person}{{Wall Street Journal}}.} \bibinfo{year}{2019}\natexlab{}.
\newblock \bibinfo{title}{Fraudsters Used AI to Mimic CEO's Voice in Unusual Cybercrime Case}.
\newblock
\urldef\tempurl%
\url{https://www.wsj.com/articles/fraudsters-use-ai-to-mimic-ceos-voice-in-unusual-cybercrime-case-11567157402}
\showURL{%
\tempurl}
\newblock
\shownote{Accessed: 23-Nov-2023}.


\bibitem[Wang et~al\mbox{.}(2023a)]%
        {valle-microsoft}
\bibfield{author}{\bibinfo{person}{Chengyi Wang}, \bibinfo{person}{Sanyuan Chen}, \bibinfo{person}{Yu Wu}, \bibinfo{person}{Ziqiang Zhang}, \bibinfo{person}{Long Zhou}, \bibinfo{person}{Shujie Liu}, \bibinfo{person}{Zhuo Chen}, \bibinfo{person}{Yanqing Liu}, \bibinfo{person}{Huaming Wang}, \bibinfo{person}{Jinyu Li}, {et~al\mbox{.}}} \bibinfo{year}{2023}\natexlab{a}.
\newblock \showarticletitle{Neural codec language models are zero-shot text to speech synthesizers}.
\newblock \bibinfo{journal}{\emph{arXiv preprint arXiv:2301.02111}} (\bibinfo{year}{2023}).
\newblock


\bibitem[Wang et~al\mbox{.}(2023b)]%
        {10243636}
\bibfield{author}{\bibinfo{person}{Zhichao Wang}, \bibinfo{person}{Xinsheng Wang}, \bibinfo{person}{Qicong Xie}, \bibinfo{person}{Tao Li}, \bibinfo{person}{Lei Xie}, \bibinfo{person}{Qiao Tian}, {and} \bibinfo{person}{Yuping Wang}.} \bibinfo{year}{2023}\natexlab{b}.
\newblock \showarticletitle{MSM-VC: High-Fidelity Source Style Transfer for Non-Parallel Voice Conversion by Multi-Scale Style Modeling}.
\newblock \bibinfo{journal}{\emph{IEEE/ACM Transactions on Audio, Speech, and Language Processing}}  \bibinfo{volume}{31} (\bibinfo{year}{2023}), \bibinfo{pages}{3883--3895}.
\newblock
\href{https://doi.org/10.1109/TASLP.2023.3313414}{doi:\nolinkurl{10.1109/TASLP.2023.3313414}}


\bibitem[Wu et~al\mbox{.}(2022)]%
        {wu22f_interspeech}
\bibfield{author}{\bibinfo{person}{Yihan Wu}, \bibinfo{person}{Xu Tan}, \bibinfo{person}{Bohan Li}, \bibinfo{person}{Lei He}, \bibinfo{person}{Sheng Zhao}, \bibinfo{person}{Ruihua Song}, \bibinfo{person}{Tao Qin}, {and} \bibinfo{person}{Tie-Yan Liu}.} \bibinfo{year}{2022}\natexlab{}.
\newblock \showarticletitle{{AdaSpeech 4: Adaptive Text to Speech in Zero-Shot Scenarios}}. In \bibinfo{booktitle}{\emph{Proc. Interspeech 2022}}. \bibinfo{pages}{2568--2572}.
\newblock
\href{https://doi.org/10.21437/Interspeech.2022-901}{doi:\nolinkurl{10.21437/Interspeech.2022-901}}


\bibitem[Yamagishi et~al\mbox{.}(2019)]%
        {vctk}
\bibfield{author}{\bibinfo{person}{Junichi Yamagishi}, \bibinfo{person}{Christophe Veaux}, \bibinfo{person}{Kirsten MacDonald}, {et~al\mbox{.}}} \bibinfo{year}{2019}\natexlab{}.
\newblock \showarticletitle{Cstr vctk corpus: English multi-speaker corpus for cstr voice cloning toolkit (version 0.92)}.
\newblock \bibinfo{journal}{\emph{University of Edinburgh. The Centre for Speech Technology Research (CSTR)}} (\bibinfo{year}{2019}).
\newblock


\bibitem[Yasur et~al\mbox{.}(2023)]%
        {yasur2023deepfake}
\bibfield{author}{\bibinfo{person}{Lior Yasur}, \bibinfo{person}{Guy Frankovits}, \bibinfo{person}{Fred~M. Grabovski}, {and} \bibinfo{person}{Yisroel Mirsky}.} \bibinfo{year}{2023}\natexlab{}.
\newblock \showarticletitle{Deepfake CAPTCHA: A Method for Preventing Fake Calls}. In \bibinfo{booktitle}{\emph{Proceedings of the 2023 ACM Asia Conference on Computer and Communications Security}} (Melbourne, VIC, Australia) \emph{(\bibinfo{series}{ASIA CCS '23})}. \bibinfo{publisher}{Association for Computing Machinery}, \bibinfo{address}{New York, NY, USA}, \bibinfo{pages}{608–622}.
\newblock
\showISBNx{9798400700989}
\href{https://doi.org/10.1145/3579856.3595801}{doi:\nolinkurl{10.1145/3579856.3595801}}


\bibitem[Yi et~al\mbox{.}(2023)]%
        {yi2023audio}
\bibfield{author}{\bibinfo{person}{Jiangyan Yi}, \bibinfo{person}{Chenglong Wang}, \bibinfo{person}{Jianhua Tao}, \bibinfo{person}{Xiaohui Zhang}, \bibinfo{person}{Chu~Yuan Zhang}, {and} \bibinfo{person}{Yan Zhao}.} \bibinfo{year}{2023}\natexlab{}.
\newblock \showarticletitle{Audio Deepfake Detection: A Survey}.
\newblock \bibinfo{journal}{\emph{arXiv preprint arXiv:2308.14970}} (\bibinfo{year}{2023}).
\newblock


\bibitem[Zhang et~al\mbox{.}(2022)]%
        {Zhang2022}
\bibfield{author}{\bibinfo{person}{Linghan Zhang}, \bibinfo{person}{Sheng Tan}, \bibinfo{person}{Yingying Chen}, {and} \bibinfo{person}{Jie Yang}.} \bibinfo{year}{2022}\natexlab{}.
\newblock \showarticletitle{A Phoneme Localization Based Liveness Detection for Text-independent Speaker Verification}.
\newblock \bibinfo{journal}{\emph{IEEE Transactions on Mobile Computing}} (\bibinfo{year}{2022}), \bibinfo{pages}{1--14}.
\newblock
\showISSN{1536-1233}
\href{https://doi.org/10.1109/TMC.2022.3187432}{doi:\nolinkurl{10.1109/TMC.2022.3187432}}


\bibitem[Zhou et~al\mbox{.}(2021)]%
        {zhou2021seen}
\bibfield{author}{\bibinfo{person}{Kun Zhou}, \bibinfo{person}{Berrak Sisman}, \bibinfo{person}{Rui Liu}, {and} \bibinfo{person}{Haizhou Li}.} \bibinfo{year}{2021}\natexlab{}.
\newblock \showarticletitle{Seen and unseen emotional style transfer for voice conversion with a new emotional speech dataset}. In \bibinfo{booktitle}{\emph{ICASSP 2021-2021 IEEE International Conference on Acoustics, Speech and Signal Processing (ICASSP)}}. IEEE, \bibinfo{pages}{920--924}.
\newblock


\end{thebibliography}

\renewcommand\thetable{\thesection.\arabic{table}}
\setcounter{table}
{0}
\renewcommand\thefigure{\thesection.\arabic{figure}}
\setcounter{figure}
{0}

\renewcommand{\thesection}{A}

\section*{Supplementary Material for \textsc{Pitch}}
\subsection*{Glossary:}

\begin{enumerate}[label=A\arabic*.]
    \item Human Cognitive Bias and Susceptibility against Deepfakes.
    \item Instructions provided to participants for data collection.
    \item Deployment Considerations.
    \item Training of Automated Challenge Compliance Detectors.
    \item Instruments for Human Evaluation of Deepfakes.
\end{enumerate}

\subsection{Cognitive Bias and Human Susceptibility} In the realm of deepfakes, individuals are often left to navigate a complex landscape independently, leaving them vulnerable to scammers who exploit their trust during phone calls. Two critical factors contribute to the rapid trust-building phenomenon in telephonic communications:

\vspace{0.35em}
\noindent\textbf{Auditory Perception Fallibility.} Humans generally acknowledge the fallibility of their auditory perception. Mishearing is a common experience, arguably more prevalent than visual misinterpretation. We naturally integrate multi-sensory inputs into listening, such as live transcription, facial expressions, mouth movements, and conversational context, to enhance our ability to discern speech. These cues are sufficiently potent to augment and even override auditory perception, as demonstrated by the McGurk effect~\cite{mcgurk1976hearing}. This effect illustrates how visual information from observing a speaker's mouth movements can alter the perceived auditory signal, leading to misheard syllables. This phenomenon underscores our innate understanding that auditory signals can be unreliable, an effect that may be amplified during audio calls where the channel is noisy and visual cues are absent. 

\vspace{0.35em}
\noindent\textbf{Technological Conditioning.} The evolving technological landscape has significantly shaped our expectations of sound quality. Individuals accustomed to the limitations of early mobile telephony -- including restricted bandwidth, dropped calls, background noise, and the need for frequent verbal confirmations -- have been conditioned to interpret and understand communication even under suboptimal auditory conditions. Despite improvements in audio clarity, exposure to imperfect sound has ingrained a tolerance for noisy or unclear speech. This tolerance accentuates our inherent bias to trust and make sense of noisy speech.

This evolutionary and technological adaptation in auditory perception is evident in how humans rate audio quality. When deploying Mean Opinion Scores (MOS) for speech naturalness assessment, the scale typically ranges from 1 (incomprehensible) to 5 (comparable to face-to-face interaction). However, the average difference between the highest and lowest MOS scores is often only two points and skewed towards the higher end~\cite{rosenberg2017bias}. This pattern indicates a degree of leniency or an adaptive bias in our auditory judgments, underscoring our propensity to make sense of auditory inputs despite variations in quality.

The innate human tendency to trust rather than mistrust auditory inputs offers a potent avenue for exploitation in social engineering attacks employing audio deepfakes.

\subsection{Data Collection Instructions}
\label{sec:instructions}

We requested each consenting participant to perform speech tasks (Table~\ref{tab:sentences}) based on the following tasks.

\begin{enumerate}[label=\arabic*., nosep, start=0]
\item  Speak the sentence normally.

\textbf{Vocal Distortions}
\item  Keep their mouth and lips static, and speak the sentence.
\item  Cup your mouth and Speak the sentence.
\item  Whisper the sentence.

\textbf{Waveform Manipulation}
\item  Speak the sentence loudly.
\item  Speak the sentence softly.
\item  Read out the sentence quickly.
\item  Read out the sentence slowly.
\item  Speak the sentence in a high pitch.
\item  Speak the sentence in a low pitch.
\item  Hold their nose and read the sentence.

\textbf{Language and Articulation}
\item  Speak a transliteration of a foreign sentence.
\item  Speak the sentence with a known accent, such as Russian or British.
\item  Sing the sentence.

\textbf{Tone of Voice}
\item  Speak the sentence with happiness or excitement.
\item  Speak the question.

\textbf{Non-Lingual Noises}
\item  Cough or Whistle.
\item  Speak the sentence while clapping.

\textbf{Playback Challenges}: \\(done while playing audio on a smartphone)
\item  Speak the sentence clearly amidst cross-talk; while playing recording of a matching sentence.
\item  Speak the sentence while playing instrumental music.
\item  Speak the sentence while playing lyrical music.
\end{enumerate}

Desktop-users attempted all challenges, while smartphone users attempted everything except playback challenges (in our naive implementation, it requires another device to play the sound). Out of 20 listed tasks we found 10 that proved valuable against deepfakes. 
\begin{table}[h]
    \centering
    \caption{The phonetically-diverse scripts used for recording data across all tasks, except for questions and foreign sentences. Questions and Foreign sentences used special variants. Expectantly participants have hard time speaking any or all foreign languages.}
    {\footnotesize\begin{tabular}{c}
\toprule
\rowcolor{gray!20} \textbf{General Sentences} \\
\midrule
The quick brown fox jumps over the lazy dog. \\
With measured words, she explained the strange rhythm of the universe. \\
Bright violets grow under the dense canopy, shading their vivid colors. \\
Jazz and classical music often feature complex harmonies and rhythms. \\
Six whimsical wizards make powerful, rhythmic chants deep in the forest. \\
The doctor's prescription advised taking unique, holistic measures. \\
The swift cheetah is an elegant and extraordinary creature. \\
The rugged path wound around the steep hill to the quaint village. \\
Gazing at the azure sky in July, the dreamy poet wrote verses. \\
The glum monk thought deeply about life's complex queries. \\
\midrule
\rowcolor{gray!20} \textbf{Questions} \\
\midrule
Does the quick brown fox jump over the lazy dog? \\
Can she explain the strange rhythm of the universe with measured words? \\
Do bright violets grow under the dense canopy, shading their vivid colors? \\
Do jazz and classical music often feature complex harmonies and rhythms? \\
Do six whimsical wizards make powerful, rhythmic chants deep in the forest? \\
Did the doctor's prescription advise taking unique, holistic measures? \\
Is the swift cheetah an elegant and extraordinary creature? \\
Does the rugged path wind around the steep hill to the quaint village? \\
While gazing at the azure sky in July, did the dreamy poet write verses? \\
Did the glum monk think deeply about life's complex queries? \\
\midrule
\rowcolor{gray!20} \textbf{Foreign Sentences} \\
\midrule
Finnish: 'En ymmarra mitaan mita tarkoitat' \\
French: 'Le coeur a ses raisons que la raison ne connait point.' \\
Spanish: 'El rapido zorro marron salta sobre el perro perezoso.' \\
German: 'Fuchse lieben es, schnell uber die grauen Hunde zu springen.' \\
Chinese: 'Feng yu zhi hou, bi jian cai hong.' \\
Russian: 'Na dvore trava, na trave drova.' \\
Arabic: 'Al-hayatu jamilatun ala al-raghmi min kulli shay'. \\
Hindi: 'Bandar bhagate hue chhat par chadh gaya.' \\
Japanese: 'Sakura no hana ga haru no kaze ni mau.' \\
Swahili: 'Simba anapenda kupumzika chini ya mti mkubwa.' \\
\bottomrule
\end{tabular}

}
    \label{tab:sentences}
\end{table}

\subsection{Practical Design Recommendations} To ensure a practical implementation of the `Deepfake-Likely' tagging system, we offer the following guidelines:

\vspace{0.35em}
\noindent\textbf{On the caller's side:}
\begin{itemize}[nosep,leftmargin=*]
\item Utilize single, succinct sentences for tasks.
\item Enhance compliance with high-quality instructional samples.
\item Avoid challenges that confuse both machines and humans, such as those involving human cognition tests or deception.
\item For playback, employ an automated sound for ease of use.
\end{itemize}
\vspace{0.35em}
\noindent\textbf{On the receiver's side:}
\begin{itemize}[nosep,leftmargin=*]
\item Employ two categories of tags: "Deepfake-Likely" and "Deepfake-Certainly," to indicate machine confidence.
\item Providing a rationale to back prediction has nominal benefits.
\end{itemize}

\subsection{Challenge Compliance Detection Architecture}
\label{sec:wav2vec2_architecture}

Our challenge compliance detection system employs a series of binary classifiers based on the Wav2Vec2 architecture~\cite{wav2vec2020}.

\vspace{0.25em}
\noindent\textbf{Wav2Vec2 Model Architecture.}
Wav2Vec2 is a self-supervised framework for speech representation learning that consists of a multi-layer convolutional feature encoder followed by a transformer context network. The architecture processes raw audio waveforms directly, eliminating the need for feature engineering.

For our implementation, we used the public \texttt{wav2vec2-base} model pretrained on 960 hours of LibriSpeech data, which was further pretrained on 50,000 hours of unlabeled speech from Libri-Light. This extensive pretraining enables the model to learn robust speech representations across diverse acoustic conditions.

\vspace{0.25em}
\noindent\textbf{Fine-tuning for Challenge Compliance Detection.}
We adapted the pretrained Wav2Vec2 model for challenge compliance detection through the following modifications:

\begin{itemize}[nosep,leftmargin=*]
    \item \textbf{Classification Head:} We replaced the original masked language modeling head with a binary classification layer consisting of a linear projection (768 → 256), followed by layer norm, ReLU activation, dropout (p=0.1), and a final linear layer (256 → 2).
    
    \item \textbf{Input Processing:} Audio samples were resampled to 16kHz and normalized to zero mean and unit variance. For each 4.5±1.1s challenge sample, we extracted fixed-length segments of 4 seconds from the center of the utterance, with zero-padding for shorter segments.
    
\end{itemize}

We implemented separate binary classifiers for each of the 18 challenges, with each classifier distinguishing between its specific challenge and regular speech (challenge \#0). The remaining 2 challenges (foreign words and questions) were evaluated using heuristic approaches based on speech transcription as described in Table~\ref{tab:detector_scores}.

\begin{table*}[h!]
\centering
\caption{Performance of Challenge Compliance Detectors. We present accuracies and AUROC values as percentages. Test accuracy is based on 10 held-out identities, while validation accuracy is averaged across 5 trained models during 5-fold cross-validation. We compare performance with human evaluation compliance ratings.}
\begin{tabular}{clC{1.5cm}ccccccc}
\toprule

\textbf{Index} & \textbf{Challenge} & \textbf{Human} &  \multicolumn{6}{c}{\textbf{Compliance Check using Finetuned Wav2Vec2}}  \\
&& \textbf{Baseline} & \textbf{Val Acc.} & \textbf{Test Acc.} &  \textbf{AUC} & \textbf{Precision} & \textbf{Recall} & \textbf{F1}\\
\midrule
\#1 & Static Mouth          & 85.9  & $89.8 \pm 3.1$ &  88.2  &  96.6  &  0.89  &  0.87  &  0.88 \\
\#2 & Cup mouth             & 86.9  & $92.0 \pm 2.0$ &  89.5  &  97.1  &  0.95  &  0.84  &  0.89 \\
\#3 & Whisper               & 94.9  & $92.9 \pm 2.4$ &  90.9  &  93.1  &  0.92  &  0.9  &  0.91\\
\#4 & Hold nose            & 43.4  & $ 83.6 \pm 3.6$ &  90.0  &  92.8  &  0.94  &  0.85  &  0.90 \\
\#5 & High Pitch           & 86.9  & $76.1 \pm  3.8$ &  81.8  &  91.2  &  0.95  &  0.67  &  0.79 \\
\#6 & Low Pitch            & 73.7  & $77.1 \pm  3.9$ &  77.3  &  85.9  &  0.75  &  0.83  &  0.78 \\
\#7 & Sing                 & 87.9  & $90.0 \pm 2.0$ &  90.0  &  96.8  &  0.89  &  0.91  &  0.9  \\
\#8 & Speak Loudly          & 86.9  & $76.1 \pm  3.8$  &  70.5  &  74.8  &  0.73  &  0.65  & 0.69\\
\#9 & Speak softly          & 85.9  & $80.7 \pm  3.8$ &  79.1  &  84.7  &  0.76  &  0.85  &  0.8\\
\#10 & Read quickly          & 98.0  & $86.8 \pm 2.6$  &  83.2  &  90.7  &  0.92  &  0.73  &  0.81  \\
\#11 & Read Slowly           & 89.9  & $80.2 \pm 1.1$ &  79.1  &  87.5  &  0.8  &  0.77  &  0.79  \\
\#13 & Accent               & 72.7  & $77.9 \pm 2.2$ &   79.5  &  85.9  &  0.81  &  0.77  &  0.79\\
\#14 & Emotions             & 89.9  & $86.2 \pm 3.7$ &  88.6  &  95.6  &  0.98  &  0.79  &  0.87  \\
\#16 & Cough/Whistle        & 90.9  & $96.3 \pm 1.8$ &  96.7  &  99.3  &  0.82  &  0.82  &  0.82 \\
\#17 & Clap                 & 84.7  & $92.1 \pm 2.8$ &  94.1  &  98.8  &  0.95  &  0.93  &  0.94 \\
\#18 & Cross-talk           & 91.5  & $94.1 \pm 2.5$ &  98.8  &  100.0  &  1.0  &  0.96  &  0.98 \\
\#19 & Instr. Playback      & 95.7  & $95.2 \pm 3.7$ &  98.1  &  99.6  &  0.98  &  0.96  &  0.97  \\
\#20 & Lyric Playback       & 100.0  & $95.9 \pm 2.3$ &  98.1  &  100.0  &  0.94  &  1.0  &  0.97 \\ \midrule
   & \textbf{Average (across all)}     & 86.0 &      87.5       & $83.6  \pm 8.0$  & 93.1  & 0.89 & 0.82 & 0.85 \\
\midrule\midrule
&&&  \multicolumn{5}{c}{\textbf{Heuristic-based Compliance Check}} & \\
\midrule
\#12 & Foreign Words   & 99.0 & \multicolumn{5}{c}{WER compared to the specific foreign script used.} &  \\
\#15 & Question        & 92.9 & \multicolumn{5}{c}{Transcribed sentence has a question mark} &  \\
\bottomrule
\end{tabular} \label{tab:detector_scores}
\end{table*}

\vspace{0.25em}
\noindent\textbf{Training Protocol.}
The training protocol for our challenge compliance classifiers consisted of:

\begin{itemize}[nosep,leftmargin=*]
    \item \textbf{Dataset Partitioning:} We partitioned the dataset by subject identity for 5-fold cross-validation, using 90 identities for training/validation and reserving 10 identities for a final hold-out test set. This identity-based partitioning ensures no speaker overlap between training and evaluation sets, testing the model's generalization to unseen speakers.
    
    \item \textbf{Optimization:} Models were fine-tuned using AdamW optimizer with a learning rate of 5e-5, $\beta_1=0.9$, $\beta_2=0.999$, weight decay of 0.01, and a linear warmup for the first 10\% of training steps followed by linear decay.
    
    \item \textbf{Training Configuration:} Each model was trained for 30 epochs with a batch size of 64.
    
    \item \textbf{Regularization:} We applied dropout (p=0.1) to the classification head and the attention layers, along with layer drop (p=0.05) for transformer layers to prevent overfitting.
    
    \item \textbf{Model Selection:} After each epoch, we monitored validation accuracy and retained the best-performing model checkpoint based on this metric.
\end{itemize}

\vspace{0.25em}
\noindent\textbf{Implementation Performance.}
Our Wav2Vec2-based challenge compliance detection system achieved:

\begin{itemize}[nosep,leftmargin=*]
    \item Mean validation accuracy of 87.5\% across all 18 chal. classifiers
    \item Mean test accuracy of 83.6\% ± 8.0\% on the hold-out test set
    \item Mean AUROC of 93.1\%, demonstrating strong discrimination
\end{itemize}

Performance metrics for individual challenge classifiers are detailed in Table~\ref{tab:detector_scores}. Notably, the system achieved the highest performance on structured audio patterns (playback challenges) and distinctive vocal characteristics (whispering, cup mouth), while performance was comparatively lower on challenges involving subtle modulations in amplitude and speaking rate.

This architecture significantly outperformed traditional approaches based on hand-crafted features such as MFCC, LFCC, or Mel spectrogram representations, validating our choice of the Wav2Vec2 framework as the backbone for challenge compliance detection.

\begin{figure*}[h!]
\centering
\includegraphics[width=1.8\columnwidth]{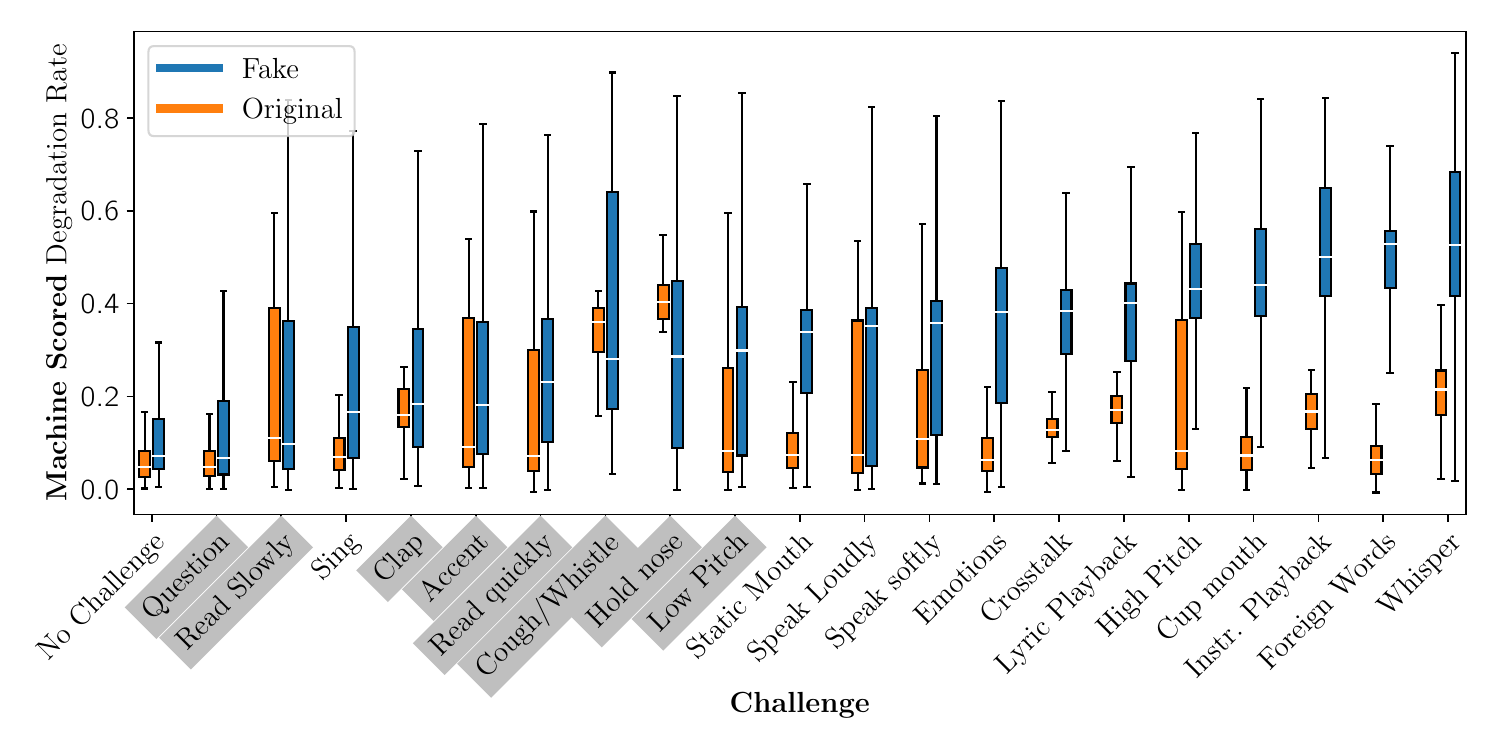}
\caption{Machine-scored degradation for fake and original audio samples across all challenges. Boxplots are arranged in ascending order based on the median values of fake samples (white bars), with outliers omitted. Higher $\uparrow$ scores indicate greater degradation and thus better challenge performance. Grayed challenges underperformed (corresponding to data in Tab~\ref{tab:automated_evaluation}).}
\label{fig:human_boxplot2}
\end{figure*}

\begin{figure*}[h!]
\centering
\includegraphics[width=1.6\columnwidth]{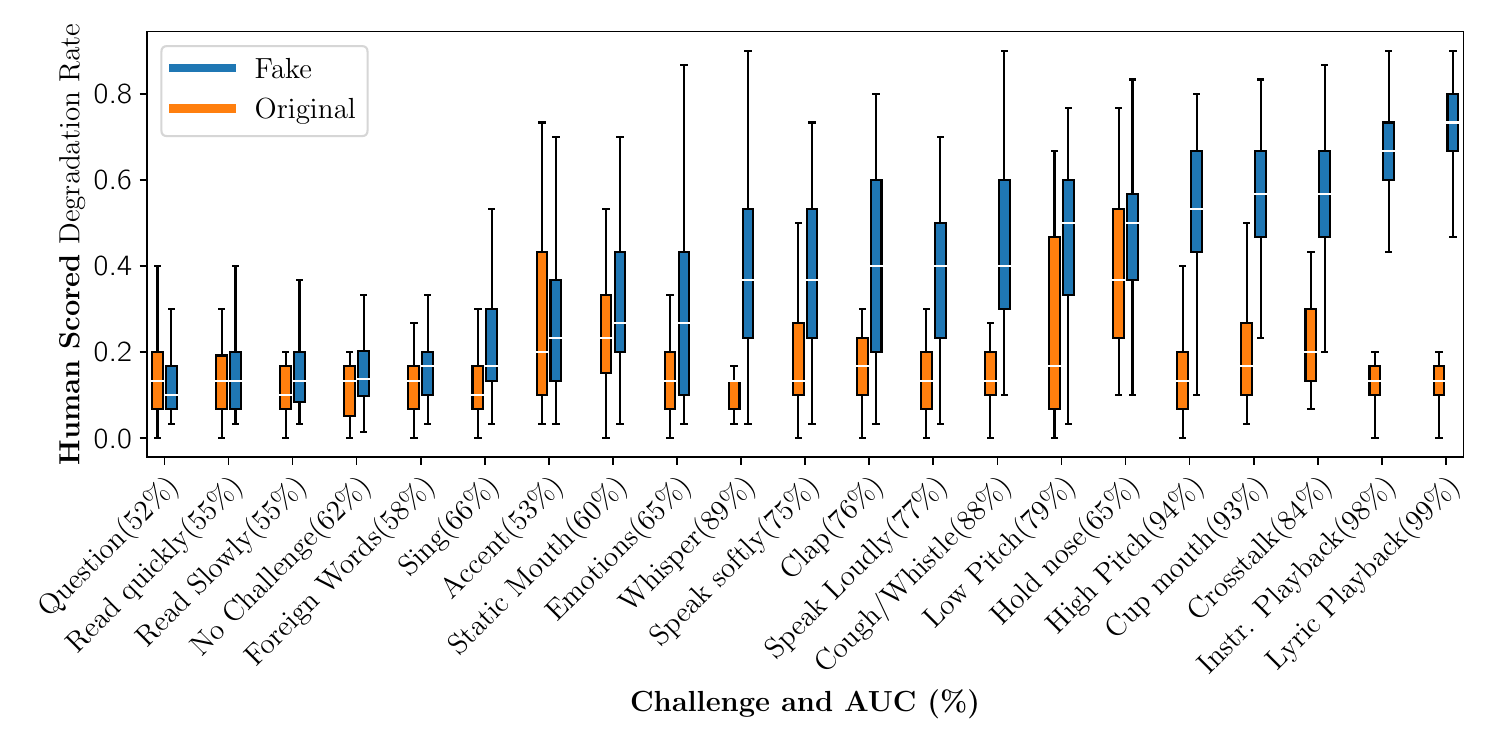}
\caption{Human-evaluated degradation scores across all challenges. Boxplots are ordered by the median of fake samples (white bars), with outliers removed. Parentheses show corresponding AUC values. Mean AUC across 20 challenges is 74.7\%.}
\label{fig:human_boxplot}
\end{figure*}

\begin{figure}
    \centering
    \includegraphics[width=0.7\columnwidth]{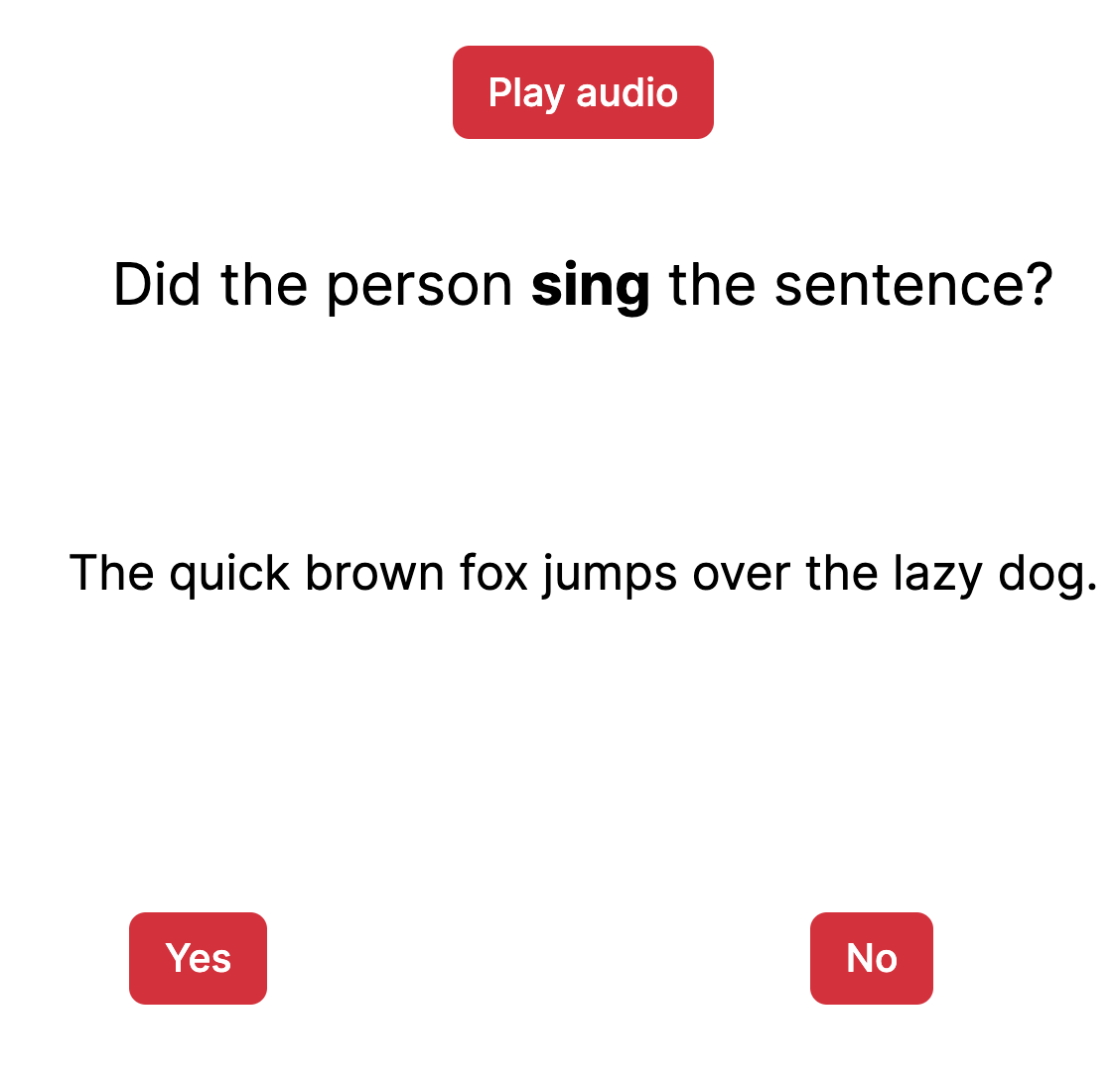}
    \caption{Human Verification of Audio Samples: Compliance verification interface.}
    \label{fig:compliance_verification}
\end{figure}

\begin{figure}
    \centering
    \fbox{\includegraphics[width=0.7\columnwidth]{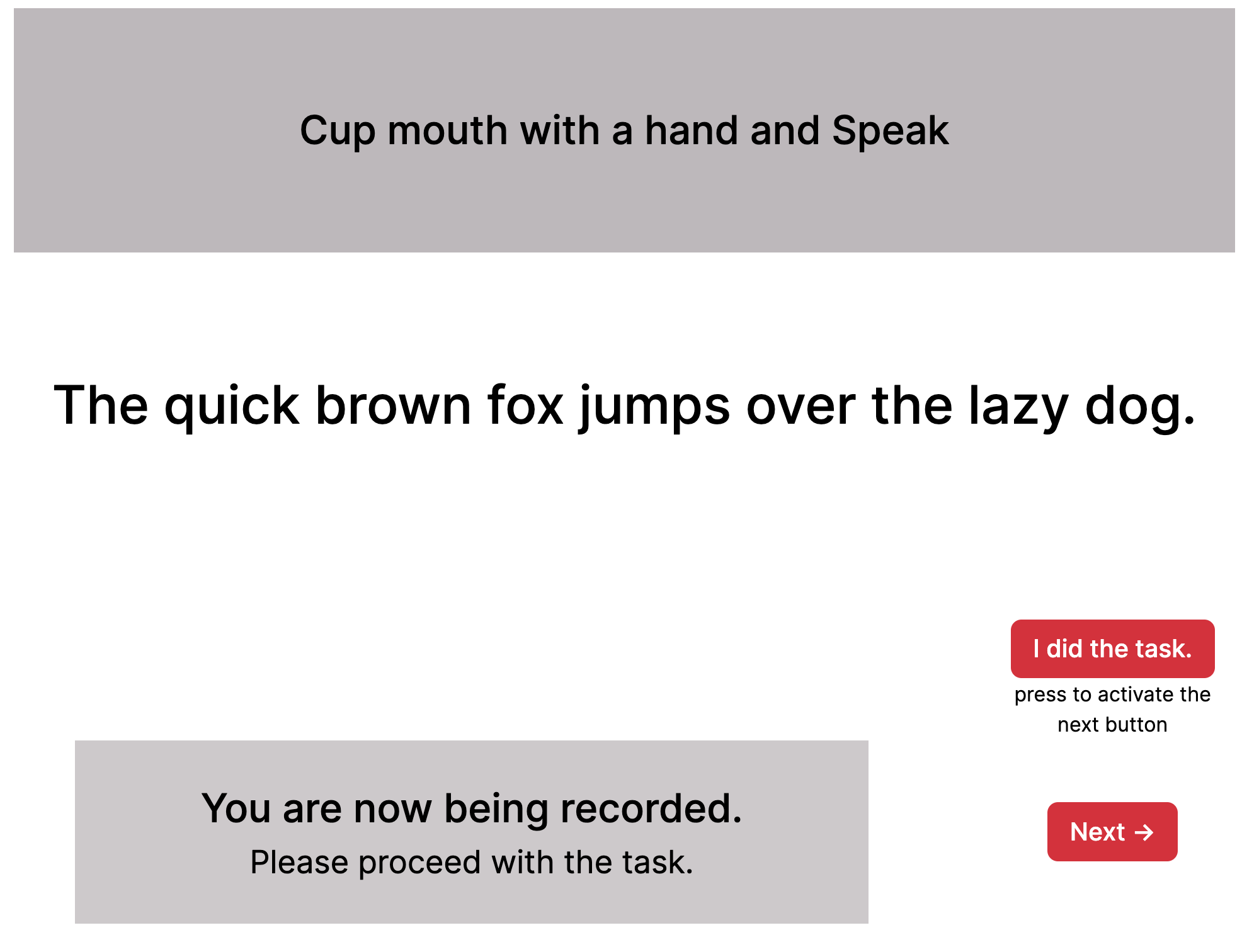}}
    \caption{Instruction screen during data collection.}
    \label{fig:data_collection}
\end{figure}

\begin{table}[t!]
\caption{Comparative performance of detection methods and human evaluation metrics across challenges.}

{\footnotesize
\begin{tabular}{L{1.5cm}C{0.7cm}C{0.7cm}|C{0.8cm}C{0.6cm}|C{0.8cm}C{0.9cm}}
\toprule
\textbf{Challenge} & \textbf{LCNN} & \textbf{SpecR-Net} &  \textbf{Star-GANv2-VC} & \textbf{PPG-VC} & \textbf{NISQA-Human $\rho$}& \textbf{Human-Human $\tau$} \\
\midrule
No Challenge       & 72.0             & 85.0         & 65.1           & 62.3           & 0.17              & 0.22              \\
Static Mouth       & 64.2             & 88.5         & 92.5           & 86.9           & 0.19              & 0.84              \\
Cup mouth          & 67.5             & 88.2         & 99.5           & 99.3           & -0.0              & 0.93              \\
Whisper            & 84.8             & 90.1         & 90.7           & 95.9           & -0.12             & 0.97              \\
Speak softly       & 70.5             & 82.6         & 75.5           & 79.3           & 0.09              & 0.79              \\
High Pitch         & 73.8             & 81.9         & 90.3           & 82.8           & -0.01             & 0.90              \\
Foreign Words      & 68.0             & 89.8         & 95.9           & 98.0           & 0.06              & 1.00              \\
Emotions           & 78.9             & 82.7         & 81.0           & 80.1           & 0.22              & 0.90              \\
Cross-talk         & 82.5             & 83.2         & 88.8           & 92.2           & -0.41             & 0.93              \\
Instr. Playback    & 80.9             & 88.2         & 97.9           & 99.2          & -0.29             & 1.00              \\
Lyric Playback     & 83.3             & 85.9         & 92.7           & 95.3           & -0.4              & 1.00              \\ \hline
Average            & 75.1             & 86.0         & 88.2           & 88.3             & -0.05             & 0.86    \\

\bottomrule
\end{tabular}
 }
\label{tab:rebuttal_table}
\end{table}

\newpage
\noindent\textbf{\textsc{Pitch} Comparison, Generalization and Human Agreement.} Table~\ref{tab:rebuttal_table} evaluates \textsc{Pitch}'s detection performance across multiple approaches and voice conversion systems. The first two columns show AUC scores for traditional deepfake detectors using LFCC features with LCNN~\cite{todisco2019asvspoof} classifier (75.1\% average) and SpecRNet~\cite{kawa2022specrnet} (86.0\% average). The middle columns demonstrate that our challenge-based approach maintains high detection performance against different voice converters: StarGANv2-VC~\cite{starganv2vc} (88.2\% AUC) and PPG-VC~\cite{ppgvc} (88.3\% AUC), comparable to our primary FREE-VC~\cite{li2023freevc} evaluation. The rightmost columns present correlation analysis between automated quality assessment (NISQA) and human ratings, revealing weak correlation ($\rho = -0.05$ average), which \textit{indicates that machines and humans assess different speech quality aspects}. This finding supports our human-AI collaborative design, as the two evaluation modalities provide complementary information. The strong human-human agreement ($\tau = 0.86$ average using Kendall's coefficient) validates the consistency and reliability of human evaluators in our studies.

\end{document}